\documentclass[aps,prb,reprint,showkeys,superscriptaddress]{revtex4-2}
\usepackage{subcaption}
\usepackage{bm,graphicx,tabularx,array,booktabs,dcolumn,xcolor,microtype,multirow,amscd,amsmath,amssymb,amsfonts,physics,siunitx}
\usepackage[version=4]{mhchem}
\usepackage[utf8]{inputenc}
\usepackage[T1]{fontenc}
\usepackage{txfonts}
\usepackage[normalem]{ulem}
\usepackage{mleftright}
\usepackage{mathtools}

\DeclareUnicodeCharacter{2212}{-}

\newcommand*{\binteg}[3]{\int_{\mathrlap{#2}}^{\mathrlap{#3}}\dd#1\:}

\usepackage[
	colorlinks=true,
    citecolor=blue,
    linkcolor=blue,
    filecolor=blue,      
    urlcolor=blue,	
    breaklinks=true
	]{hyperref}
\DeclareMathOperator{\erfc}{erfc}

\newcommand{\SupMat}{\hyperlink{SI}{supplementary material}}

\newcommand{\ie}{\textit{i.e.}\ }
\newcommand{\etal}{\textit{et al.}\ }

\newcommand{\VU}{Department of Chemistry \& Pharmaceutical Sciences and Amsterdam Institute of Molecular and Life Sciences (AIMMS), Faculty of Science, Vrije Universiteit, 1081HV Amsterdam, The Netherlands}
\newcommand{\UCM}{Department of Chemistry and Biochemistry, University of California Merced, 5200 North Lake Rd. Merced, CA 95343, USA}
\newcommand{\FSU}{Department of Mathematics, Florida State University, Tallahassee, FL 32306-4510, United States of America}
\begin{document}	

\title{Real space Mott--Anderson electron localization with long-range interactions: exact and approximate descriptions}

\author{Antoine \surname{Marie}}
\affiliation{\VU}
\author{Derk P. \surname{Kooi}}
\affiliation{\VU}
\author{Juri \surname{Grossi}}
\affiliation{\UCM}
\author{Michael \surname{Seidl}}
\affiliation{\VU}
\author{Ziad H. \surname{Musslimani}}
\affiliation{\FSU}
\author{Klaas \surname{Giesbertz}}
\affiliation{\VU}
\author{Paola \surname{Gori-Giorgi}}
\affiliation{\VU}

\begin{abstract}
Real materials always contain, to some extent, randomness in the form of defects or irregularities.
It is known since the seminal work of Anderson that randomness can drive a metallic phase to an insulating one,
and the mechanism responsible for this transition is intrinsically different from the one of the interaction-induced transitions discovered by Mott.
Lattice Hamiltonians, with their conceptual and computational advantages, permitted to investigate broadly the interplay of both mechanisms.
However, a clear understanding of the differences (or not) with their real-space counterparts is lacking, especially in the presence of long-range Coulomb interactions.
This work aims at shedding light on this challenging question by investigating a real-space one-dimensional model of interacting electrons in the presence of a disordered potential.
The transition between delocalized and localized phases are characterized using two different indicators, namely the single-particle occupation entropy and the position-space information entropy.
In addition, the performance of density functional approximations to reproduce the exact ground-state densities of this many-body localization model are gauged.
\end{abstract}


\maketitle

%=================================================================%
\section{Introduction}
\label{sec:intro}
%=================================================================%

Being able to describe the metallic or insulating behavior of any given material is certainly a desirable feature for an electronic structure method.
Unfortunately, this question is quite nettlesome and remains an open problem.
One of the challenges is that various intrinsically different mechanisms can drive a metal to an insulating phase and would need to be described at the same time.
For example, in Mott insulators the transition is due to electron-electron interactions \cite{Mott_1949,Mott_1990}.
Indeed, if interactions become predominant over kinetic energy this will lead to the localization in space of the electronic density.
Hence, this decrease of the electrons mobility induces a decrease of the conductivity of the material.
These interaction-driven metal-insulator transitions are not restricted to long-range Coulomb interactions, as Hubbard showed that Mott transitions can also occur in lattice model Hamiltonians with short-range on-site interactions \cite{Hubbard_1963}.
On the other hand, systems  of non-interacting particles can also be driven to the insulating phase but through a totally different mechanism.
The seminal work of Anderson showed that the conductivity of a system of non-interacting particles can go to zero as soon as the external medium/potential is disordered enough \cite{Anderson_1958}.
Of course, actual electrons are interacting and any real material has a given degree of disorder, so a clear understanding of the interplay of these two phenomena is an important challenge towards the description of real materials.
Note that other mechanisms of metal-insulator transitions exist but are out of the scope of this work \cite{Imada_1998,Hasan_2010}.

One of the challenges when studying such phenomena is to find a way to characterize the different phases as well as transitions between them.
The usual approach to distinguish metals and insulators is to consider their low-energy excitation spectra \cite{Byczuk_2005,Shinaoka_2009}.
However, in a seminal paper, Walter Kohn showed that the arrangement of electrons in the many-body ground state already contains all the information necessary to distinguish between metallic and insulating phases. \cite{Kohn_1964}
A plethora of indicators trying to quantify the (de)localization of electrons have been designed and investigated since.
In the following we will mention some of them but this list shall not be considered exhaustive.

As the previously mentioned Hubbard model, lattice models in general have been a crucial tool to study Mott and Anderson transitions as well as their interplay. \cite{Shinaoka_2009,Canella_2021,Canella_2022}
One of the key advantages of lattice Hamiltonians is that the associated Hilbert space has a tensor-product structure which gives a straightforward way of evaluating entanglement. \cite{Larsson_2006,Wu_2006}
Single-site entanglement measures are a natural way of quantifying correlation in lattice models and as such are a powerful tool to identify transition between different phases.
It has been applied to study the Mott transition in various dimensions, \cite{Gu_2004, Larsson_2005, Coe_2010, Coe_2011} Anderson transitions,\cite{Canella_2021}  Mott--Anderson transitions \cite{Franca_2008,Canella_2021,Canella_2022} and more generally many-body localization in spin chains and various other models. \cite{Kjall_2014,Luitz_2015}

For continuous space Hamiltonians, it is \textit{a priori} not straightforward to define such entanglement measures, so other quantities have been designed to study metal-insulator transitions from the ground-state perspective.
Note that these quantities can be, and have been, also used in lattice models.
Concepts imported from information theory like the Shannon information entropy can be useful to study the localization of electrons.
If the electronic density is considered as a continuous random variable, then the associated Shannon information entropy will quantify its lack of information.
This means that it will be maximal for a totally delocalized density and minimal if the electrons are fully localized. \cite{Amovilli_2004,Coe_2008}
Other information entropy like the Kullblack-Leibler divergence applied to densities corresponding to different orbitals can also be useful. \cite{Luitz_2015}
Using the two-body reduced density matrix (RDM) in addition to the electronic density to compute the localization tensor can give further valuable information about the arrangement of the electrons within the ground-state, hence about their metallic or insulating character. \cite{Resta_2006, Angyan_2009, Bendazzoli_2010, KeralaVarma_2015, Diaz-Marquez_2018}
In addition to the electronic density or the two-body RDM, the single-electron picture provided by the eigenfunctions and eigenvalues of the one-body RDM (natural orbitals and their occupation numbers), is also an available tool to study such transitions.
The natural occupation numbers can be used to compute the single-particle occupation entropy, also known as von Neumann entropy of the one-body RDM, correlation or Jaynes entropy. \cite{Amovilli_2004, Coe_2008, Bera_2015, Diaz-Marquez_2018, Wang_2021}
The natural orbitals give complementary information about the system, \ie its characteristic localization length, through the inverse participation ratio. \cite{Bera_2015}

Besides understanding and characterising the physics of Mott--Anderson transition, it is crucial to know which computationally affordable approximate methods are able to capture it. 
One of the most successful methods to compute ground-state energies and densities, both in quantum chemistry and condensed matter physics, is Kohn-Sham (KS) density functional theory (DFT). \cite{Kohn_1965}
Although exact in principle, KS DFT must in practice rely on approximations for the exchange-correlation (XC) functional, which typically struggle to describe strongly correlated systems.
The well-known local density approximation (LDA) gives a qualitatively good description of metallic electrons but fails badly to describe the insulating electrons in the Mott phase, and so do all other current semilocal and hybrid (mixing Hartree-Fock exchange) approximations \cite{YinBroLopVarGorLor-PRB-16}.
A promising way of describing strong correlation within DFT is based on an expansion in the limit of infinite coupling strength, \cite{Seidl_1999,Seidl_2007,Gori-Giorgi_2009,Lewin_2018,Cotar_2018} yielding the so-called strictly correlated electrons (SCE) functional. \cite{Malet_2012,Malet_2013,Malet_2014,Mendl_2014}
The resulting KS SCE formalism, although far from perfect, has been proven successful in qualitatively describing systems in which the electron-electron interaction is predominant over the kinetic energy, \ie Mott insulators like systems.
One can then wonder if this approximation is still sufficient to describe strongly interacting particles in a disordered medium, shedding light to new avenues to build XC approximations. \cite{Karlsson_2018}

The aim of this work is precisely to investigate these two crucial points, namely i) the characterisation of the Mott--Anderson physics for real-space Hamiltonians with long-range interactions, and ii) its description within KS DFT.
To this purpose, we have selected a setting that allows us to obtain very accurate solutions (that we call ``exact'' in the following) for the many-body Hamiltonian: few electrons confined in a one-dimensional box with a disordered potential, interacting with the Coulomb long-range potential (renormalised at contact to mimic interactions in a thin quantum wire \cite{Bednarek_2003}). We first address point i) by investigating the Mott--Anderson physics in the exact case, studying different indicators and their ability to characterise the different regimes, moving to point ii) by studying the performance of various approximations (LDA, SCE, and also exact exchange), compared to the exact case. We analyse densities and XC potentials, providing insights to build new XC functionals. 

The paper is organised as follows: in the following Sec.~\ref{sec:theoretical}, the system that will be studied is described in detail, and a brief outline of KS SCE theory is given, while computational details are provided in Sec.~\ref{sec:comp_det}.
The interaction- and disorder-induced transitions are investigated separately in Sec.~\ref{sec:mott} and Sec.~\ref{sec:anderson}.
The central point of the manuscript, which is the interplay of interaction and disorder, is divided in two parts, Sec.~\ref{sec:interplay_exact} and  Sec.~\ref{sec:interplay_approx}, in which the exact and approximate description of this interplay are described, respectively.
Finally, Sec.~\ref{sec:conclusion} draws some conclusions.

%To the best of our knowledge, investigation of the Mott--Anderson physics in the continuous case from a ground-state perspective has not been done and giving insights into this problem is one of the aim of this manuscript.
%In addition to understanding the physics of this phenomena as well as its differences (or not) with its lattice models counterparts, this manuscripts aims at investigating approximate description of the same system as of course the exact solutions can only be obtained in few electrons systems.

%=================================================================%
\section{Theoretical background}
\label{sec:theoretical}
%=================================================================%
\subsection{Hamiltonian}
Throughout this manuscript the following $N$-electron one-dimensional Hamiltonian will be considered
\begin{equation}
    \label{eq:hamiltonian}
    \hat{H} = - \frac{1}{2} \sum_i^N \pdv[2]{x_i} + \sum_{i < j} w_{\text{int}}(\abs{x_i - x_j}) + \sum_i v_{\text{ext}} (x_i)
\end{equation}
where the external potential is
\begin{equation}
    \label{eq:external_potential}
    v_{\text{ext}} (x) = v_{\text{box}} (x) + v_{\text{rand}} (x)
\end{equation}
with
\begin{align}
  \label{eq:box_potential}
   v_{\text{box}}(x) =
  \begin{cases}
    0& -\frac{L}{2} \leqslant x \leqslant \frac{L}{2} \\
    +\infty& x \in \mathbb{R}\setminus [-\frac{L}{2}, \frac{L}{2}]\\
  \end{cases}
\end{align}
and
\begin{equation}
  \label{eq:random_potential}
  v_{\text{rand}}(x,V,\sigma) = V \sum_{i=1}^M v_i e^{-\frac{(x-X_i)^2}{2\sigma^2}},
\end{equation}
\ie the random potential is constituted of $M$ Gaussians centered at random positions $X_i$ with width $\sigma$ and amplitudes $v_i V$ where $v_i$ are random numbers between 0 and 1. If we remove the interparticle interaction, the parameter $L$ in the remaining single-particle Hamiltonian \eqref{eq:hamiltonian} can be absorbed by scaling, such that $x \in [-\frac{1}{2}, \frac{1}{2}]$. With the interparticle interaction present in Eq.~\eqref{eq:hamiltonian}, the parameter $L$ becomes an effective interaction-strength parameter.
Hereafter we stick to these scaled units, and fix $M=300$ and $\sigma= 0.02$. With this choice, by increasing $V$ we can go from no disorder, to the quantum tunneling regime ($E\ll V\ll \sigma^{-2}$), up to trivial localization in just one of the random potential wells.
We will refer to a given couple of random sets $\{v_i\}$ and $\{X_i\}$ as a \textit{realization} of disorder.
Within a realization, two parameters can be varied, the disorder strength $V$ and the interaction strength $L$.

In one-dimensional systems, the short-range divergence of the Coulomb potential $1/|x|$ would force nodes in the wave function whenever two particles are at a contact, and would make the mean-field Hartree functional of Eq.~\eqref{eq:Hartree_funct} infinite. The physics of three-dimensional systems with Coulomb interaction is then much better mimicked by an interaction that is finite at contact.
In this work the interaction function $w_{\text{int}}$ in \eqref{eq:hamiltonian} is chosen to be
\begin{equation}
  \label{eq:interaction}
  w_{\text{int}}(x) = \frac{\sqrt{\pi}}{2b}\exp(\frac{x^2}{4b})\erfc\biggl(\frac{x}{2b}\biggr),
\end{equation}
which is the effective electron-electron interaction corresponding to a model 1D quantum wire of thickness $b$, in which the lateral degrees of freedom have been averaged over a narrow harmonic confinement. This interaction still behaves as $1/|x|$ for large $x$. \cite{Bednarek_2003}
Hereafter $b$ is set to 0.1.
Note that an LDA parametrization is available for this interaction. \cite{Casula_2006,Abedinpour_2007}

\subsection{Density Functional Theory and its approximations}
It is known, thanks to the Hohenberg--Kohn theorem, \cite{Hohenberg_1964} that the ground-state energy $E_0$ of the Hamiltonian \eqref{eq:hamiltonian} is a functional of the electronic density $\rho$,
\begin{align}
    \label{eq:energy}
    E_0 &= \min_\rho E[\rho] \\
    &= \min_\rho \mleft\{ \min_{\Psi \to \rho} \mel{\Psi}{\hat{T} + \hat{W}_{\text{int}}}{\Psi} + \int \dd \vb{r}~v_{\text{ext}}(\vb{r}) \rho(\vb{r}) \mright\} \notag
\end{align}
where $\hat{T}$ and $\hat{W}_{\text{int}}$ are the kinetic energy and electron-electron interaction operators, respectively.
The first term of the right hand side in \eqref{eq:energy} is the so-called Levy--Lieb functional denoted hereafter as $F[\rho]$. \cite{Levy_1979}
In the KS formulation of DFT, \cite{Kohn_1965} this latter is expressed as
\begin{equation}
  \label{eq:KS_funct}
  F[\rho] = T_s[\rho] + U[\rho] + E_{\text{xc}}[\rho],
\end{equation}
the different terms being the KS kinetic energy
\begin{equation}
  \label{eq:KS_kinetic}
  T_s[\rho] = \min_{\Psi \to \rho} \mel{\Psi}{\hat{T}}{\Psi},
\end{equation}
the Hartree energy
\begin{equation}
  \label{eq:Hartree_funct}
  U[\rho] = \frac{1}{2} \int \dd x_1\dd x_2~w_{\text{int}}(\abs{x_1 - x_2})\rho(x_1)\rho(x_2),
\end{equation}
and the remainder $E_{\text{xc}}[\rho]$, known as the exchange-correlation energy functional, contains all the complexity of the Levy--Lieb functional and needs to be approximated in practice.

Unfortunately, designing approximations for $E_{\text{xc}}[\rho]$ which successfully describe strongly correlated systems is a nettlesome problem.
In particular, the functionals built with the traditional ingredients forming the so-called Jacob's ladder of DFT are known to fail in such situations.
On the other hand, the strictly correlated electrons (SCE) formalism utilizes intrinsically different ingredients which can be used to construct functionals adapted to strong correlation.
In the KS-SCE scheme the Hartree-exchange-correlation functional $E_{\text{Hxc}} = U + E_{\text{xc}}$ is approximated as \cite{Malet_2012} 
\begin{equation}
  \label{eq:SCE_functional}
  W_{\text{int}}^{\text{SCE}}[\rho] = \min_{\Psi \to \rho} \mel{\Psi}{\hat{W}_{\text{int}}}{\Psi},
\end{equation}
\ie as the minimum of the electronic interaction over all the wave functions yielding the density $\rho$.
Note the close connection between Eq.~\eqref{eq:SCE_functional} and the KS kinetic energy functional of Eq.~\eqref{eq:KS_kinetic}.
In other words, the KS-SCE approximation replace the minimum of the sum in the Levy--Lieb functional as the sum of the minima
\begin{equation}
    \label{eq:KSSCEfunc}
    F[\rho] = \min_{\Psi \to \rho} \mel{\Psi}{\hat{T} + \hat{W}_{\text{int}}}{\Psi} \approx T_s[\rho] + W_{\text{int}}^{\text{SCE}}[\rho],
\end{equation}
thus providing a lower bound to $F[\rho]$.

While the KS functional defines a fictitious system of non-interacting particles with density $\rho$, the SCE functional defines in analogous way a fictitious system of infinitely interacting particles yielding the density $\rho$. 
In the SCE system the position of an electron determines the position of the remaining $N-1$ electrons, therefore describing the situation of perfect correlation. 
The position of the electron $i$ is quantified by the co-motion function $f_i[\rho](x)$ where $x$ is the position of the first electron. These co-motion functions (or ``optimal maps'') can be constructed exactly~\cite{Sei-PRA-07,ColDepDim-CJM-15} for the one-dimensional case considered here.
The probability of finding the first electron at $x$ should be equivalent to the one of finding the electron $i$ at $x_i$ so the co-motion functions must fulfill the following condition
\begin{equation}
    \label{eq:prob_comotion}
    \rho(x) \dd x = \rho\bigl(f_i(x)\bigr) \dd f_i(x).
\end{equation}
In addition, the indistinguishability of the electrons is most naturally imposed by enforcing the following group structure on the co-motion functions \cite{Seidl_2007}
\begin{align}
        \label{eq:group_pty}
        f_1(x) &= x, \;
        f_2(x) = f(x),\;
        \dotsc, \;
        f_N(x) = f^{N-1}(x) \quad \text{and} \notag \\
        f^N(x) &= x, \;\text{where} \;
        f^N(x) = \underbrace{f \circ f \circ \dots \circ f}_{N~\text{times}}(x). 
\end{align}
The SCE functional is fully determined by the co-motion functions $f_i(x)$, \cite{SeiGorSav-PRA-07,MirSeiGor-JCTC-12}
\begin{equation}
    \label{eq:SCE_analytical}
    W_{\text{int}}^{\text{SCE}}[\rho] = \frac{1}{2} \int \dd x~\rho(x) \sum_{i=2}^N w_{\text{int}}(\abs{x - f_i(x)}),
\end{equation}
and its functional derivative with respect to the density $\rho(x)$, $v^{\text{SCE}}[\rho](x)\equiv \delta W_{\text{int}}^{\text{SCE}}[\rho]/\delta\rho(x)$ satisfies the equation \cite{Malet_2013}
\begin{equation}
    \label{eq:SCE_potential}
    \pdv{v^{\text{SCE}}[\rho](x)}{x} = \sum_{i=2}^N w'_{\text{int}}(\abs{x - f_i(x)})~\text{sign}(x - f_i(x)).
\end{equation}
Therefore, from the co-motion functions corresponding to a given density $\rho$, one can compute the SCE functional \eqref{eq:SCE_analytical} and the SCE potential by integration of \eqref{eq:SCE_potential}. In one-dimensional systems, the co-motion functions are exactly\cite{Sei-PRA-07,ColDepDim-CJM-15} obtained by integration of \eqref{eq:prob_comotion} with the total suppression of fluctuations boundary conditions
\begin{equation}
    \label{eq:bound_cond}
    \binteg{x'}{f_i(x)}{f_{i+1}(x)} \rho(x') = 1,
\end{equation}
which yields the following expression
\begin{align}
  \label{eq:1D_comotion}
  f_i(x) =
  \begin{cases}
    N_e^{-1}\bigl(N_e(x) + i - 1\bigr)  &x \leq N_e^{-1}(N+1-i),  \\
    N_e^{-1}\bigl(N_e(x) + i - 1 - N\bigr) &x > N_e^{-1}(N+1-i),\\
  \end{cases}
\end{align}
where $N_e(x)$ is the electronic density cumulant function
\begin{equation}
    \label{eq:cumulant}
    N_e(x) = \binteg{x}{\infty}{x}\rho(x).
\end{equation}
The KS equations can then be solved self-consistently using this analytical expression of the co-motion functions to compute the SCE potential, which acts as an approximation to the Hartree and exchange-correlation potential. This approximation is asymptotically exact in the low-density limit.
In addition to this SCE formalism, two more typical KS approximations will be considered for comparison, namely LDA and exact exchange (EXX).
The LDA parametrization for this one-dimensional interaction is taken from Ref.~\onlinecite{Casula_2006,Abedinpour_2007}. 
The EXX exchange-correlation functional has only the exchange part, which is taken to be the KS one. 
For the case of two electrons, the KS exchange functional is equal to minus the half of the Hartree functional defined in~\eqref{eq:Hartree_funct}, and for the ground state it is equivalent to the Hartree--Fock approximation.

\subsection{Localization indicators}
Before turning to the results of this manuscript, we discuss the different indicators that will be used and gauged throughout this work.
The main indicator considered here is the single particle occupation entropy
\begin{equation}
    \label{eq:vonNeumann1RDM}
    S = -\Tr(\gamma\ln(\gamma)) = - \sum_\alpha n_\alpha \ln(n_\alpha).
\end{equation}
 where $n_\alpha$ are the natural spin-orbital occupation numbers of $\gamma$, with
\begin{multline}
    \label{eq:}
    \gamma(x,x') = \int \Psi(x,x_2,\dotsc,x_N)^* \\
    \Psi(x',x_2,\dotsc,x_N)~\dd x_2\dotsi\dd x_N ,
\end{multline}
the one-body RDM associated to the wave function $\Psi$.
Bera \etal showed that $S$ can be used, or more precisely its variance with respect to various realizations of disorder, as an indicator of disorder-induced localisation transition in the Anderson--Hubbard model. \cite{Bera_2015}
Note that this is not specific to this model as it is now well-known that strong fluctuations are ubiquitous at the edge of the many-body localization transition. \cite{Kjall_2014,Luitz_2015,Vosk_2015}
This entropy is also referred as correlation entropy, as $S$ increases when the correlation in the system increases. \cite{Collins_1993, Wang_2021}
This will be a useful property to study the Mott transition (see Sec.~\ref{sec:mott}).

However, this entropy has a few drawbacks for our purpose.
The sole Anderson transition in Sec.~\ref{sec:anderson} is defined for non-interacting particles, hence in this case the single-particle entropy is constant because the natural occupation numbers are constants equal to $0$ or $1$ for all disorder strengths.
In addition, because the auxiliary KS system uses non-interacting particles to approximate the true interacting system, the KS single-particle occupation entropy is constant for all interaction strengths $L$ and all disorder strengths $V$.
Therefore, the position-space information entropy will be considered as an alternative to the single-particle entropy of Eq.~\eqref{eq:vonNeumann1RDM}.
This alternative entropy is defined as the Shannon entropy of the electronic density normalized to one electron, \ie
\begin{equation}
  S_\rho= - \int \frac{\rho(x)}{N}\ln(\frac{\rho(x)}{N}) \dd x.
\end{equation}
The Shannon entropy of a continuous probability variable quantifies its lack of information.
Therefore, in this case it means that $S_\rho$ will be maximal for a uniform density and minimal if the electrons are perfectly localized.

%=================================================================%
\section{Computational details}
\label{sec:comp_det}
%=================================================================%

\subsection{Numerically accurate many-body calculations}

Near exact many-body wave functions have been obtained by representing the Hamiltonian of Eq.~\eqref{eq:hamiltonian} as a sparse matrix using a grid of 512 equidistant points, Dirichlet boundary conditions and tenth-order finite-difference approximation for the second order derivative.
Then, the diagonalization was performed using Krylov iterative subspaces as well as the filtering algorithm described in Ref.~\onlinecite{Zhou_2007}.
The convergence criteria was chosen as $\abs*{\hat{H}\Psi -E\Psi} < 10^{-10}$.
Because the computational cost of these near-exact solutions is growing exponentially fast with the number of particles, this study is restricted to two electrons.
Moreover, only singlet wave functions has been considered as the ground state is always a singlet for this system (in the large interaction limit the singlet and triplet become degenerate).

\subsection{DFT calculations}

The various flavors of KS DFT considered in this work have been implemented with the aid of the the open-source \textsc{pyscf} package. \cite{Sun_2020} To represent the kinetic energy a second-order finite-difference approximation was used on a grid of 512 equidistant points. The computation of the cumulant $N_e(x)$, the co-motion functions $f_i(x)$, the SCE functional $W^\mathrm{SCE}_\mathrm{int}[\rho]$ and its functional derivative $v^\mathrm{SCE}[\rho](x)$ was performed using \verb|jax|. \cite{jax2018github}
For large interaction strengths, level shifts has been used to converge the KS SCE solutions.

%=================================================================%
\section{Results}
\label{sec:results}
%=================================================================%

%%%%%%%%%%%%%%%%%%%%%%%%%%%%%%%%%%%%%%%%%%%%%%%%%%
\subsection{Mott transition}
\label{sec:mott}
%%%%%%%%%%%%%%%%%%%%%%%%%%%%%%%%%%%%%%%%%%%%%%%%%%

%%%  FIG 1  %%%
\begin{figure*}
    \centering
    \includegraphics[width=\linewidth]{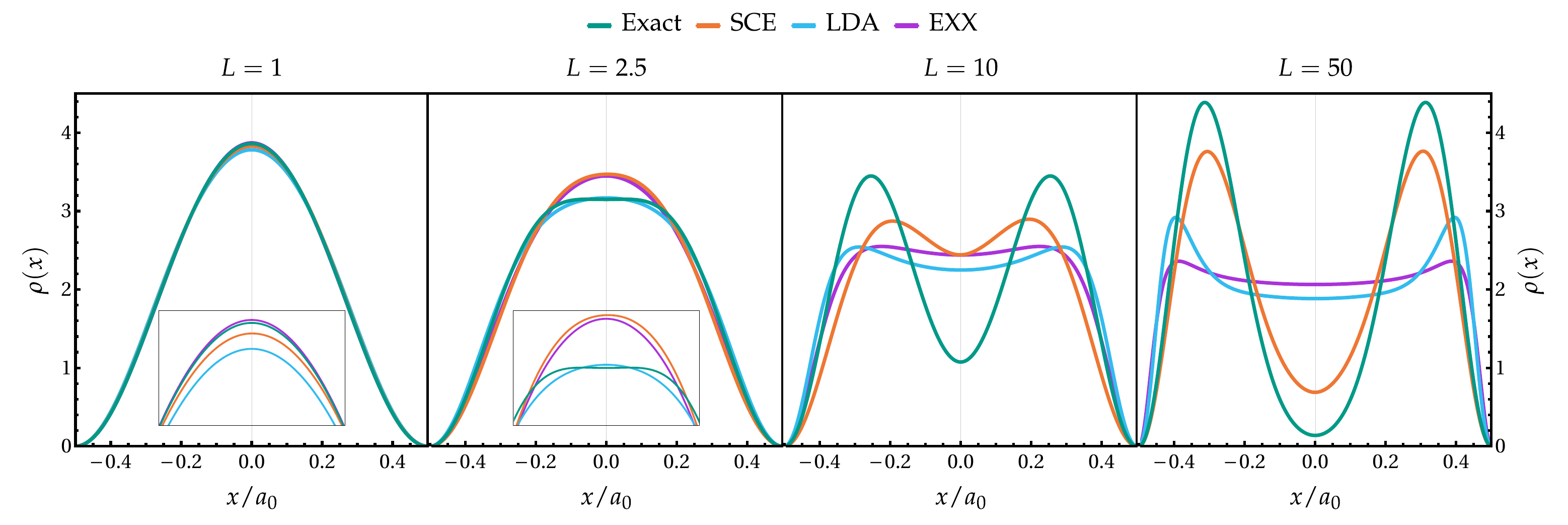}
    \caption{Exact, KS SCE, KS LDA and KS EXX ground-state electronic densities for two electrons in a box without external potential for various effective interaction strength $L$.
    \label{fig:fig_1}}
\end{figure*}
%%% %%% %%% %%%

%%%  FIG 2  %%%
\begin{figure}
    \centering
    \includegraphics[width=\linewidth]{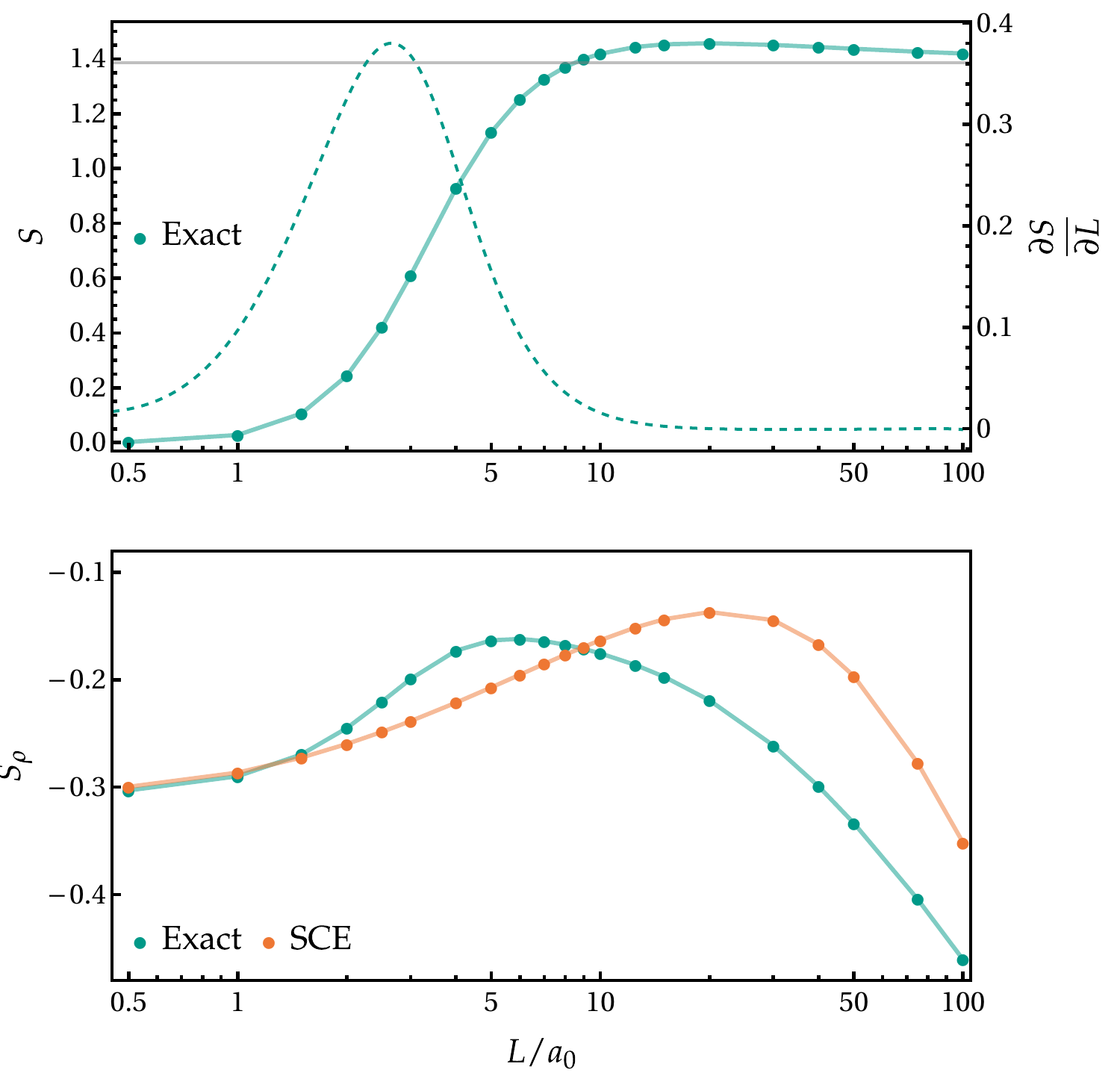}
    \caption{Single-particle occupation entropy (top panel) and position-space information entropy (bottom panel) of the ground-state of two electrons in a box without external potential for various effective interaction strength $L$. The dashed lines are the derivative of the interpolated solid lines. 
    \label{fig:fig_2}}
\end{figure}
%%% %%% %%% %%%

%%%  FIG 3  %%%
\begin{figure*}
    \centering
    \includegraphics[width=\linewidth]{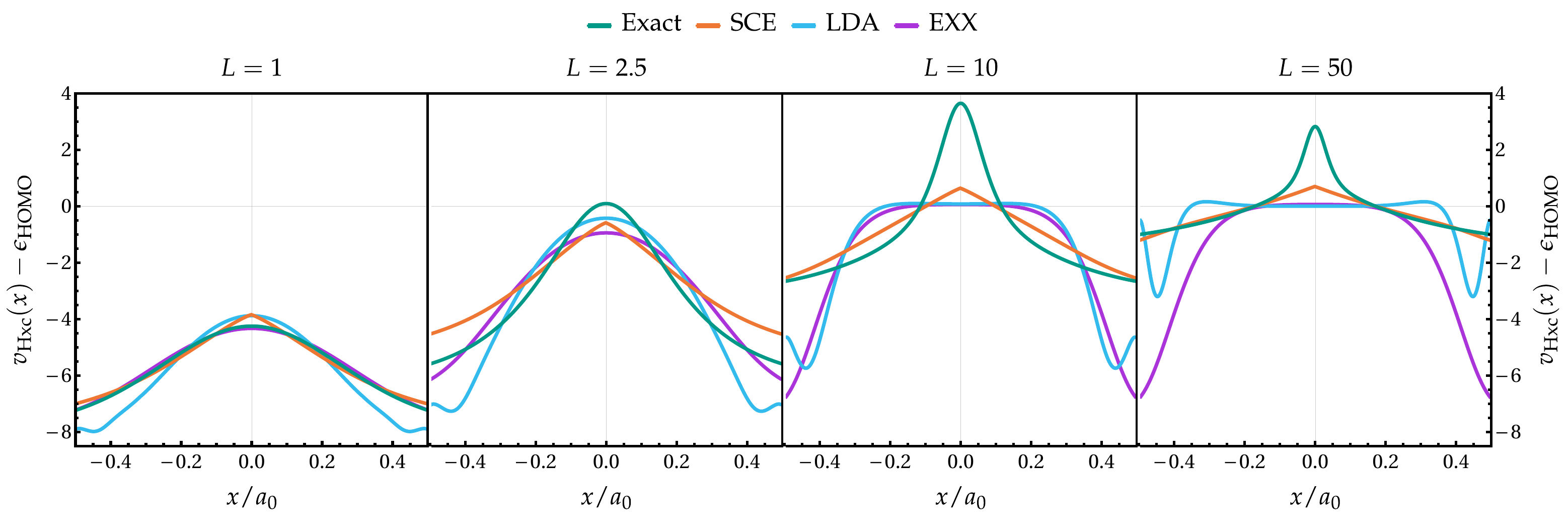}
    \caption{Exact, KS SCE, KS LDA and KS EXX ground-state electronic Hartree-exchange-correlation KS potentials (minus the energy of the highest occupied molecular orbital) for two electrons in a box without external potential for various effective interaction strength $L$.
    \label{fig:fig_3}}
\end{figure*}
%%% %%% %%% %%%

In order to focus on the sole effect of interaction, the random potential of Eq.~\eqref{eq:random_potential} is set to 0 in this subsection. 
This special case of Mott transition where the localization is only due to electron-electron interaction, \ie the positions of the localised electrons are not simply determined by the external potential, is known as a Wigner transition (see Ref.~\cite{Roy_2019}).
Figure~\ref{fig:fig_1} shows the exact ground-state electronic densities (green lines) of two interacting electrons in a box for various values of the effective interaction strength $L$.
The difference is readily seen between the two regimes in which either the kinetic energy or the interaction energy is predominant.
For $L = 1$ the density is centered around $x=0$ and the two electrons are delocalized over the whole box, while for larger interactions each electron is localized in one side of the box  (see the $L = 100$ panel).

The associated single-particle occupation entropy is given in the top panel of Fig.~\ref{fig:fig_2} (solid line).
When $L$ tends to zero, the entropy tends to zero as well, which means that in this limit the two opposite spin electrons occupy the same orbital.
As soon as the interaction is increased, $S$ starts to grow and eventually goes to an asymptotic limit value equal to $2\ln(2)$ which corresponds to two singly occupied orbitals.
The derivative of this entropy with respect to the interaction strength $L$ is displayed as well (dashed line).
Note that the derivative is slightly negative for $L \gtrsim 20$.
The maximum of the derivative ($L=2.7$) marks the onset of the Mott-like transition and can be used to define a critical Mott interaction strength.
Alternatively, one could define this $L_c$ as the value of $L$ at which the entropy reach the $2\ln(2)$ value, which gives $L_c=8.5$.
Note that the definition of $L_c$ is somewhat arbitrary in this case but this concept will be useful later to study the influence of disorder on interaction-induced transition.

The bottom panel shows the corresponding position-space information entropy.
This entropy is increasing for small values of $L$ before decreasing when $L$ goes to infinity as expected because the electrons are localized in this limit.
This maximum of the entropy can be understood by looking at the second panel of Fig.~\ref{fig:fig_1} ($L=2.5$) where one can see that the density is enlarged when compared to the leftmost panel.
Indeed, the density is deformed due to this stronger repulsion between the electrons, yet, the interaction is not strong enough to localize them on each side of the box.
Therefore, the position-space information entropy goes through a maximum of delocalization at intermediate $L$ before decreasing towards localization.
The position of this maximum can be used as a definition of $L_c$ for this indicator.
This gives a value of $L_c$ equal to $5.8$ which is in qualitative agreement with what has been observed using the single-particle occupation entropy.

The densities obtained with three different KS approximations (SCE, LDA and EXX) are also plotted in Fig.~\ref{fig:fig_1}.
In the weak interaction regime every approximation gives fairly good results in terms of the density.
Note that even if the SCE density is correct for small $L$, the associated SCE total energy is a poor approximation to the exact value in this high-density limit. \cite{Malet_2012}
When the interaction strength is increased, LDA and EXX fail to reproduce the localization of the electrons.
At $L=100$, their densities are almost totally delocalized.
LDA looks slightly better than EXX, because it can produce two small localized bumps, but this is likely a boundary effect (for example, the two bumps in LDA disappear in an harmonic confinement, while still present in the exact and SCE case \cite{Malet_2012,Malet_2013}).
On the other hand, SCE is able to localize the electrons on each side of the box, getting at least qualitatively right results. However, SCE still does not localize the electrons enough, as can be seen on the two rightmost panels of Fig.~\ref{fig:fig_1}.
The SCE functional finds the minimum of the interaction energy for a given density (see Eq.~\eqref{eq:SCE_functional}), hence this energy is underestimated with respect to the exact one, which leads to this slight under-localization in the large $L$ limit.

That this localization of electrons happens for larger interaction strengths in SCE than in the exact case can also be observed by looking at the corresponding position-space information entropies.
Indeed, according to Fig.~\ref{fig:fig_2} the value for $L_c$ in the SCE case is $21.5$.
We do not report the entropy for the two other approximate KS methods as they fail to localize the electrons therefore the associated entropies are not meaningful.
Thus, in this case the Shannon entropy is a qualitatively good indicator that can be used for KS approximations.

% %%%  TAB 1  %%%
% \begin{table}
%     \caption{Exact singlet and triplet ground-state energies of two electrons in a box without external potential. Their KS SCE and LDA counterpart are also reported}
%     \label{tab:tab_1}
%     \begin{ruledtabular}
%         \begin{tabular}{ccccc}
%             $L$ & Exact singlet & Exact triplet & KS SCE singlet & LDA singlet \\
%             \hline
%             0.5 & 45.2491 & 102.797 & 43.8945 &  \\
%             1 & 14.2219 & 27.2620 & 12.7002 &  \\
%             2.5 & 3.93820 & 5.12189 & 2.83425 &  \\
%             10 & 0.521574 & 0.524640 & 0.389929 &  \\
%             25 & 0.137692 & 0.137700 & 0.119259 &  \\
%             100 & 0.0219934 & 0.0219934 & 0.0212647 &  \\
%         \end{tabular}
%     \end{ruledtabular}
% \end{table}
% %%% %%% %%% %%%

To conclude this section focusing on the sole role of interaction, the exact KS potential (obtained by reverse engineering as described in Ref.~\onlinecite{Giarrusso_2022}) is compared to the SCE, the LDA and the EXX ones in Fig.~\ref{fig:fig_3}. We have reported the Hartree-exchange-correlation potential (which inside the box corresponds to the total KS potential) minus the highest occupied orbital energy (HOMO). We see that in the weakly correlated regime (small $L$) there are no classically forbidden regions inside the box, a case that is qualitatively well described by all approximations. The onset of interaction-induced localization corresponds to a peak in the exact KS potential that creates classically forbidden regions inside the box. This peak is known to have an important correlation kinetic energy component,\cite{BuiBaeSni-PRA-89,GriBae-PRA-96,TemMarMai-JCTC-09,Helbig_2009,HodRamGod-PRB-16,Kohut_2016,Hodgson_2017} which is missing in the SCE potential.\cite{YinBroLopVarGorLor-PRB-16} We see however that SCE does have a peak, though not enough pronounced, and it is the only approximation able to create a classically forbidden region in the center of the box. Both EXX and LDA have a completely wrong behavior for large $L$: except for boundary effects, they are converging towards a uniform density with uniform external potential. The self-consistent creation of classically forbidden regions to localise the charge density appears to be a key feature for the description of strong correlation in KS DFT.

%%%%%%%%%%%%%%%%%%%%%%%%%%%%%%%%%%%%%%%%%%%%%%%%%%
\subsection{Anderson transition}
\label{sec:anderson}
%%%%%%%%%%%%%%%%%%%%%%%%%%%%%%%%%%%%%%%%%%%%%%%%%%

%%%  FIG 4  %%%
\begin{figure}
    \centering
    \includegraphics[width=0.95\linewidth]{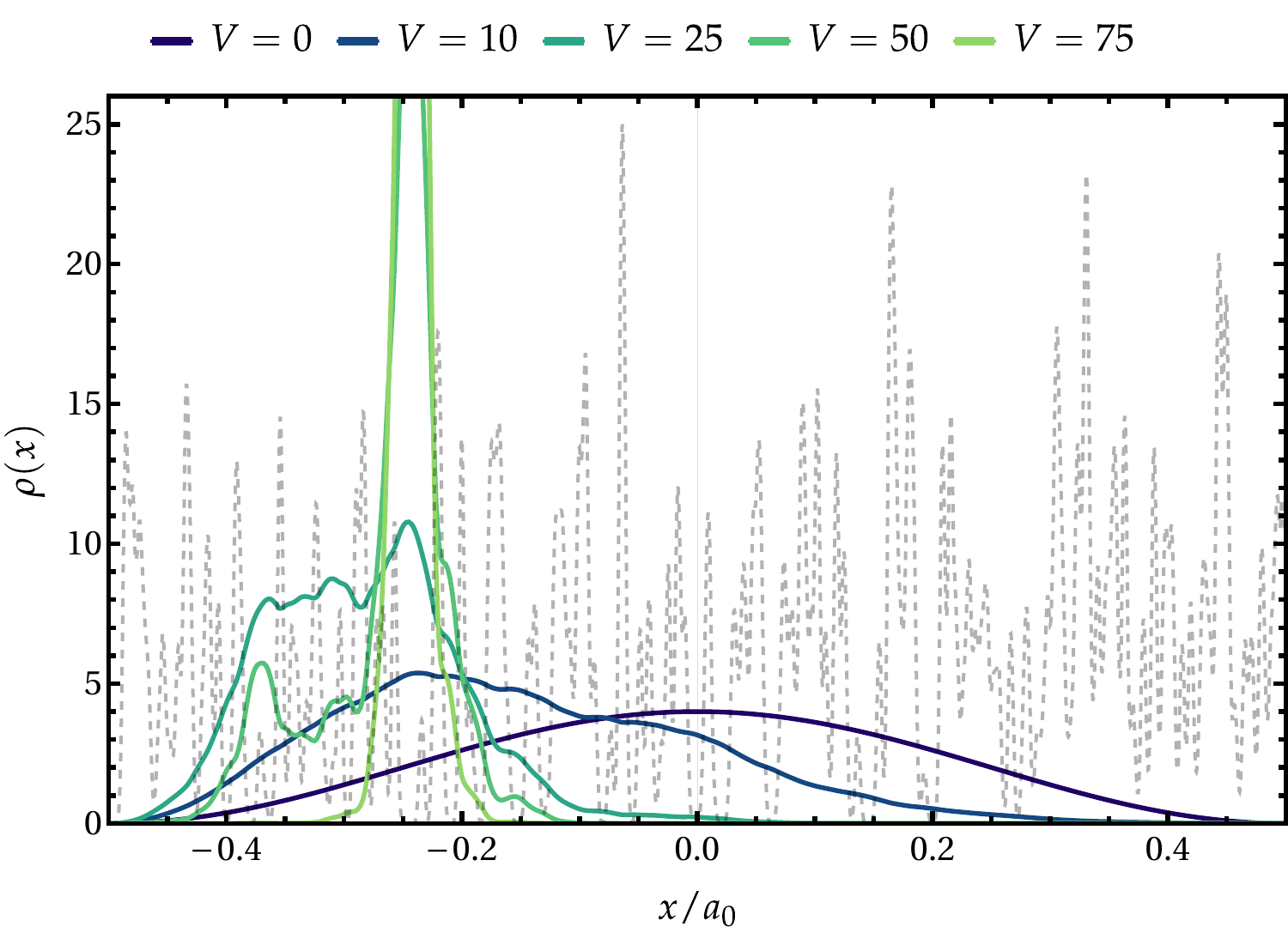}
    \caption{Electronic densities (solid lines) of two non-interacting electrons for various values of the potential strength $V$. The height of the potential (dashed line) is non-indicative, it has been adjusted just to show its shape on the same plot.
    \label{fig:fig_4}}
\end{figure}
%%% %%% %%% %%%

%%%  FIG 5  %%%
\begin{figure}
    \centering
    \includegraphics[width=0.95\linewidth]{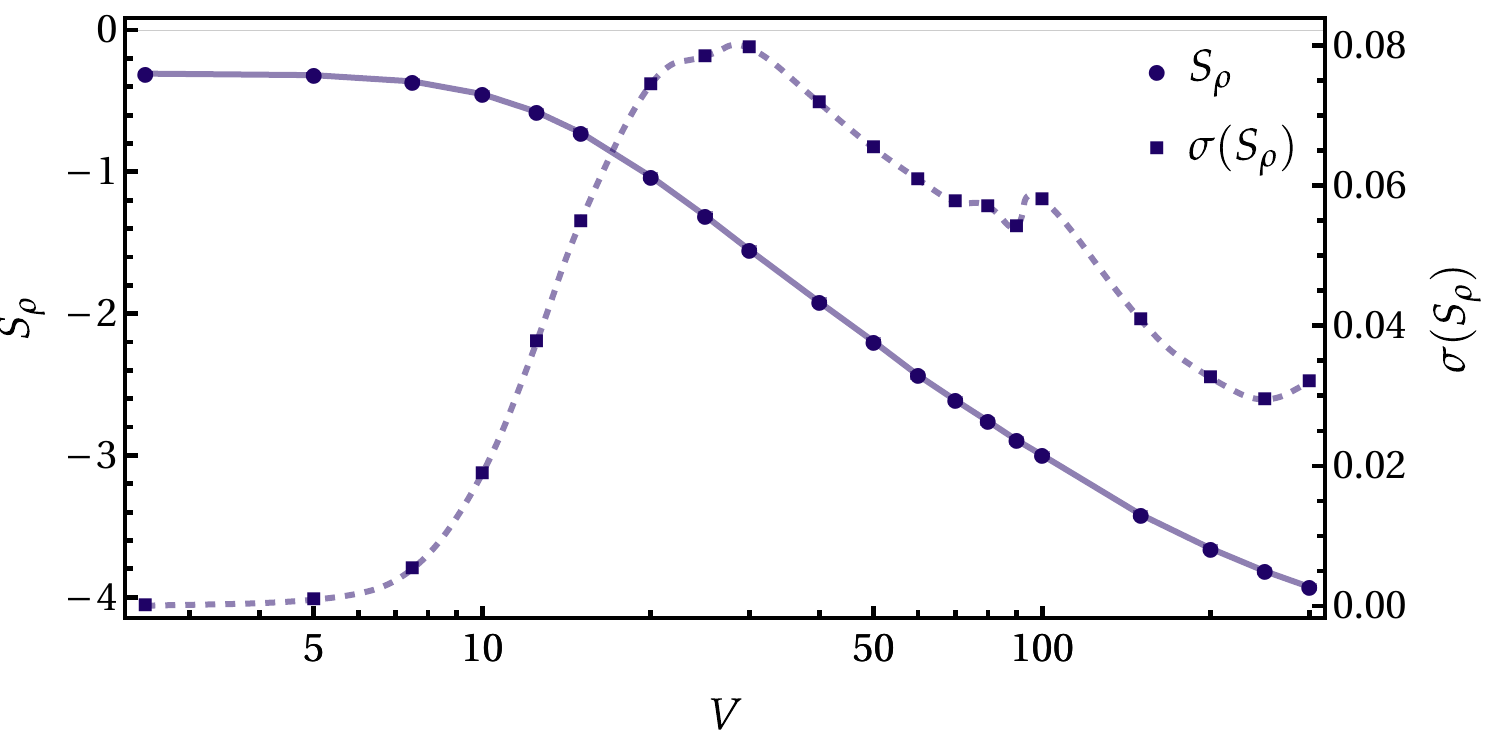}
    \caption{Mean value (solid line) and variance (dashed line) of the Shannon entropy of the densities of two non-interacting electrons by averaging over 500 different realizations of the random potential~\eqref{eq:random_potential}.
    \label{fig:fig_5}}
\end{figure}
%%% %%% %%% %%%

In this second subsection, the other physical phenomena of interest for this study is isolated, namely the influence of the disordered potential, thus removing the interaction term in Eq.~\eqref{eq:hamiltonian}.
Figure~\ref{fig:fig_4} shows the ground-state densities of two non-interacting electrons for various values of the disorder strength $V$ (see Eq.~\eqref{eq:random_potential}) corresponding to the same realization of disorder.
It is clear that the electrons become more and more localized as the disorder strength is increased.
Throughout this evolution three different regimes can be distinguished.
First, for small $V$ the density is delocalized over a large part of the box (see $V=10$).
On the other hand, for large disorder strengths the electrons are localized in (almost) only one well of the random potential. 
This regime is referred to as ``trivial localization'' as the potential hills are so large that the two electrons will simply localize in the lowest well.
For intermediate $V$, the competition between localization due to the external potential and the kinetic energy which tends to delocalize the electrons is more subtle.
For example, at $V=25$ the electrons are localized in one part of the box, yet delocalized over several wells and hills of potentials.
This is the so-called ``quantum tunnelling regime''.
This plot has been reproduced for other realizations of disorder in the \SupMat.

At this point it is interesting to note a major difference between the localized phases due to interaction or disorder.
In the former case, the electronic density has a two-peak structure with each peak integrating to one. Each electron is not pinned to one side, they can both be found on both sides but never on the same side. Hereafter, we will refer to it as electron localization for the sake of conciseness while it would be more precise to name it charge density localization. Note also that in this study we do not consider localization due to artificial symmetry breaking.
On the other hand, in the latter case the two electrons are both in the same localized orbital.
Due to this difference, the existence of a localized phase in the presence of both interactions and disorder is a subtle question.
Indeed, interactions and disorder could interfere destructively and break the localization.
On the contrary, the two mechanisms could reinforce each other and accentuate it.
Before looking at this interplay, the indicators of localization as a function of the disorder strength are analyzed.

Because the two particles are non-interacting the single-particle occupation entropy is of no use in this case.
The position-space information entropy averaged over 500 realizations for various values of disorder strength is plotted in Fig.~\ref{fig:fig_5} (circle dots).
As expected from the densities of Fig.~\ref{fig:fig_4}, the entropy is continuously decreasing when the disorder strength is increasing.
However, this smooth decrease of $S_\rho$ does not give much information on the transition between the three regimes introduced previously.
One can gain more insights on this by considering the variance of $S_\rho$.
Figure~\ref{fig:fig_5} shows the variance averaged over the 500 realizations.
Its peak structure is characteristic of the large fluctuations happening at the transition regime between the delocalized and localized phases, \ie the quantum tunnelling regime.
Then, one can define a critical disorder strength $V_c$ as the position of this peak, which gives $V_c=29.5$.

%%%%%%%%%%%%%%%%%%%%%%%%%%%%%%%%%%%%%%%%%%%%%%%%%%
\subsection{Interplay of interaction and disorder: exact solutions}
\label{sec:interplay_exact}
%%%%%%%%%%%%%%%%%%%%%%%%%%%%%%%%%%%%%%%%%%%%%%%%%%

%%%  FIG 6  %%%
\begin{figure*}
    \centering
    \includegraphics[width=\linewidth]{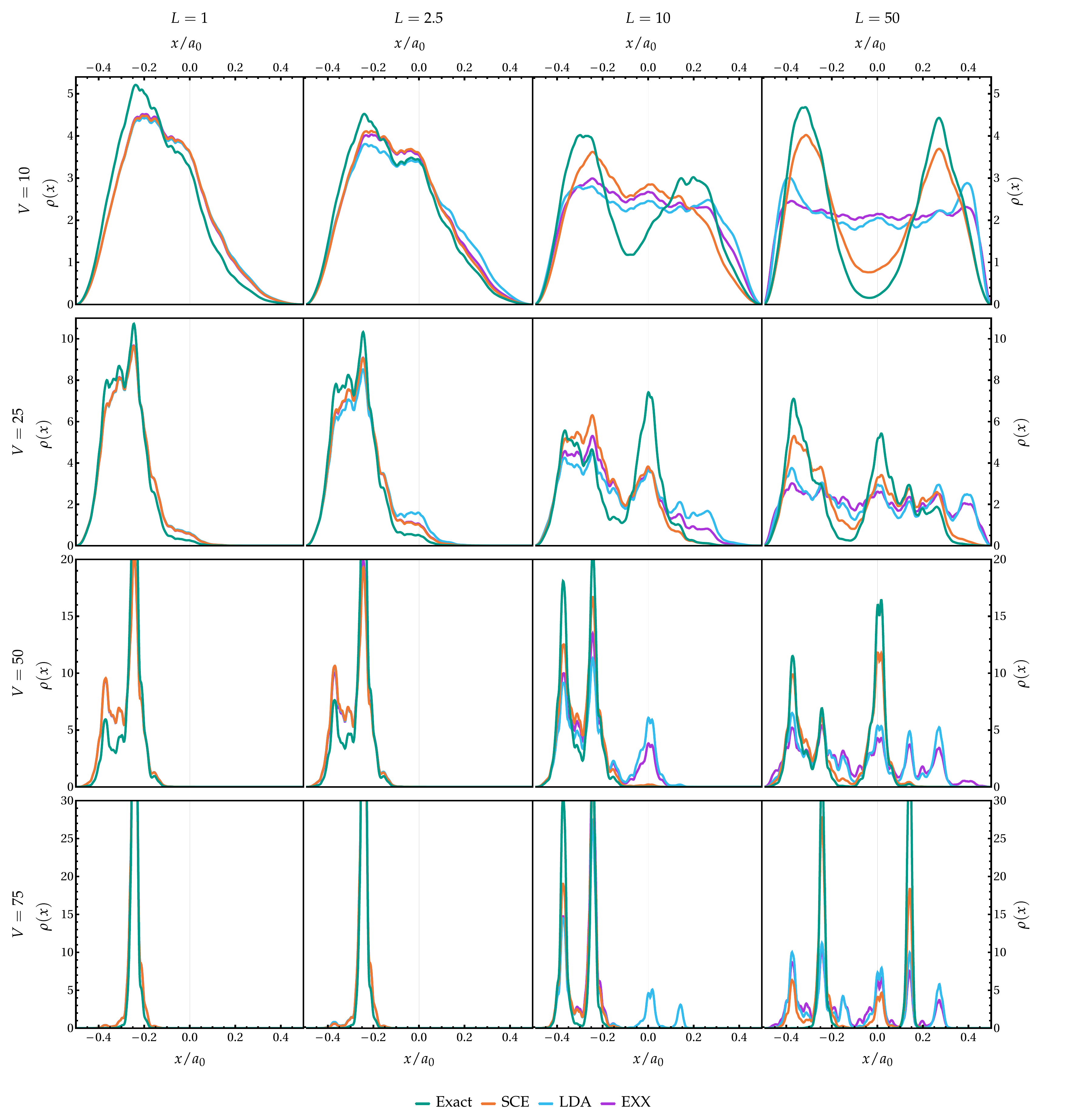}
    \caption{Exact, KS SCE, KS LDA and KS EXX ground-state electronic densities for a given realization of disorder with various effective interaction strengths $L$ and disorder strengths $V$.
    \label{fig:fig_6}}
\end{figure*}
%%% %%% %%% %%%

The interplay of interactions and disorder is now considered. 
As a first step, the focus is only on the exact solutions and the approximate descriptions will be investigated in the following subsection.
Figure~\ref{fig:fig_6} shows the ground-state densities for various values of the interaction and disorder strengths (note that this plot has been reproduced for other realizations of disorder in the \SupMat).
For this subsection, only the green curves corresponding to the exact solutions need to be considered.
Rows correspond to interaction-induced transitions while columns describe disorder-induced transitions.
The first row corresponds to an interaction-induced localization transition in the presence of weak disorder. 
In this case the random potential alters slightly the surface of the densities but the overall shapes of the density along the transition is similar to the case without disorder displayed in Fig.~\ref{fig:fig_1}.
Similarly, the disorder-induced transition in the presence of weak interactions (see first column) is very close to the Anderson transition for non-interacting particles of Sec.~\ref{sec:anderson}.

However, in the nine remaining panels, where both the interactions and the disordered potential have considerable effects, their interplay becomes more interesting.
For example, the rightmost column shows the densities for increasing disorder of two strongly interacting electrons.
One can see that each particle is going through the three regimes of the Anderson transition on their respective side of the box.
Note that the quantum tunnelling regime of the left and right particles do not happen at the same disorder strengths.
On the other hand, the Mott-like transition in the presence of a strong external potential (see last row) is much steeper than the one of Fig.~\ref{fig:fig_1}.
Indeed, in this case the density is not deformed continuously by the increase of the interaction strength. 
The two electrons stay confined in the lowest well of potential until the interaction becomes too strong so that one electron jumps into another localized orbital.
If the interaction is further increased, one of the electron will continue to jump in other localized orbitals further away from the remaining electron (see $V=50$ and $75$ for $L=50$).

%%%  FIG 7  %%%
\begin{figure}
    \centering
    \includegraphics[width=\linewidth]{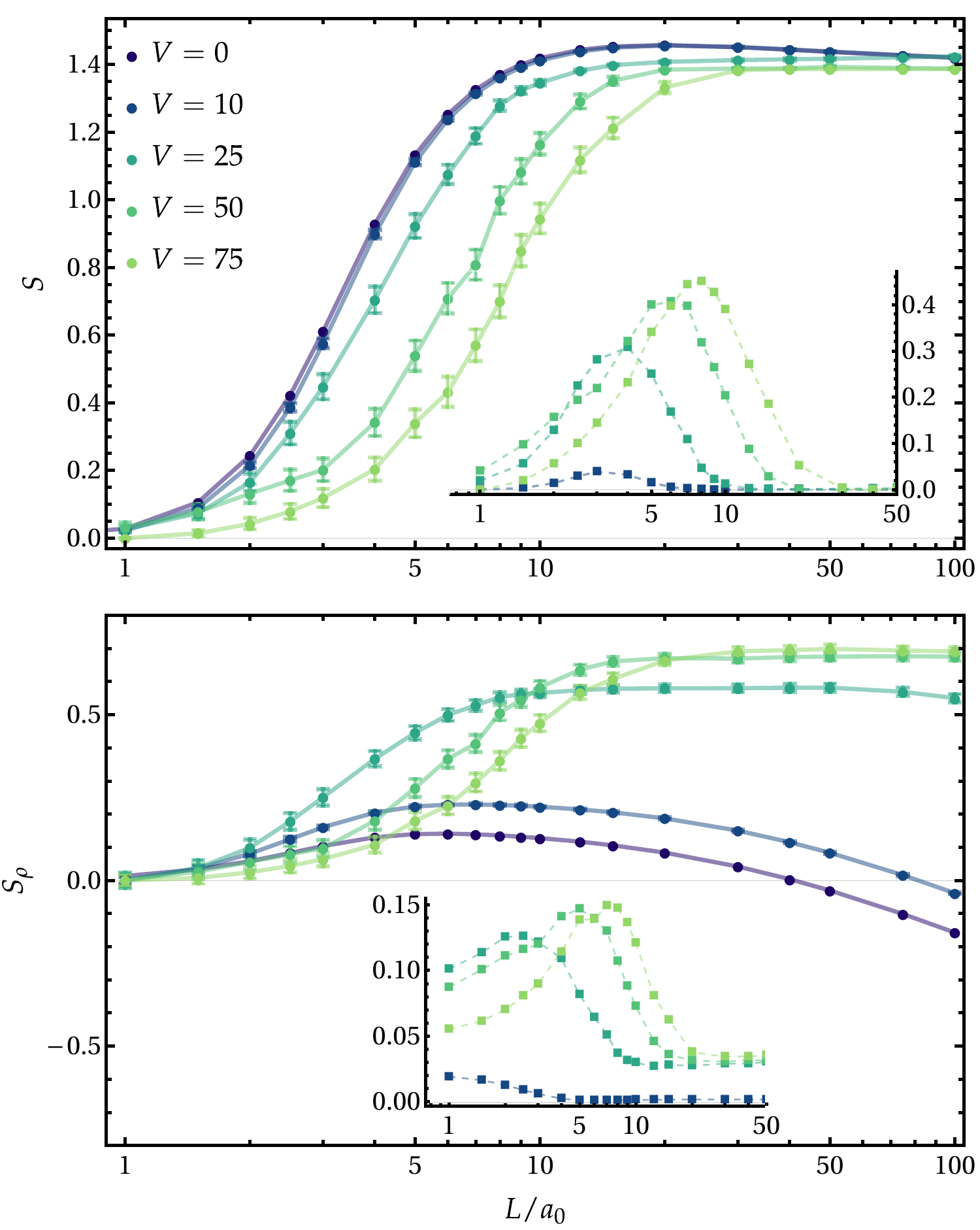}
    \caption{Single-particle occupation entropy (top panel) and position-space information entropy (bottom panel) as a function of $L$ for various disorder strengths $V$ averaged over 200 realizations. The two insets represent the variance associated to the entropies of their respective panel.
    \label{fig:fig_7}}
\end{figure}
%%% %%% %%% %%%

In between these extreme regimes, the densities exhibits a subtle mix of both localization mechanisms, \ie the disorder potential which drives the electrons in a part of the box while the interactions between electrons favors a two peaks structure.
This plot of the densities helps to grasp qualitatively the interplay of interest for this study, however to draw conclusions about it one needs to consider statistic means of indicators of localization over a large number of realizations.

Figure~\ref{fig:fig_7} shows the evolution with respect to $L$ of the means, over 200 realizations of disorder, of the single-particle occupation entropy (top panel) and the position-space information entropy (bottom panel) for various values of $V$.
It is readily seen on the top panel that the step structure of $S$ observed in Fig.~\ref{fig:fig_1} is preserved by the inclusion of the external potential.
However, the random potential leads to a delay of this jump of $S$ from $0$ to $2\ln(2)$.
This means that the Mott-like transitions are shifted to larger interaction strengths as the Coulomb interaction needs to overcome the effect of disorder which favors putting both electrons in the same localized orbital. 
In addition to the mean of the single-particle occupation entropy, also its variance is considered and plotted in the inset of the top panel of Fig.~\ref{fig:fig_7}.
Note that there is no variance for $V=0$, since all realizations are the identical.
In the presence of disorder, the variance of $S$ exhibits the expected peak structure at the delocalization-localization edge. 
The position of the maxima of the various variance plots can be used to define an alternative critical interaction strengths $L_c$. 
The ordering (and qualitatively their position) of these maxima agrees with the ones obtained using the two arbitrary definitions of $L_c$ defined in Sec.~\ref{sec:mott} for $V=0$.

Now turning to the second indicator of interest in this study, the plots of $S_\rho(L)$ for different disorder strengths are displayed at the bottom panel of Fig.~\ref{fig:fig_7}.
Note that they have been shifted to ease the visualisation of their differences, the unshifted values are plotted in the \SupMat.
The interpretation of the effect of the random potential on $S_\rho$ as a function of $L$ is not as straightforward.
Indeed, the overall shape of the curves is not the same in the weak and strong disorder regime.
For weak disorder, the observed trend is the same as for the $V=0$ case of Fig.~\ref{fig:fig_2}.
Hence, one can use the same criteria to define a critical interaction strength.
However, for medium and large disorder the $S_\rho(L)$ function has a smooth-step shape (see $v=25, 50$ and $75$).
In this regime, the densities evolve from one localized peak to two localized peaks.
Therefore, they are more delocalized and $S_\rho$ is larger in the large $L$ limit.
One can still define the critical interaction strength as the value of $L$ at which the step reach the maximum plateau.
Hence, the position-space information entropy seems to exhibit the same trend for the values of $L_c$ as a function of disorder as the single-particle occupation entropy.

To confirm this, the variance of the position-space information entropy is considered as well. 
In the medium-large disorder strength regime, the variance displays the characteristic peak structure.
However in the weak disorder case the variance seems to be of no use to locate the transition.
Therefore, the position-space information entropy may not be a good choice, or at least should be used with care, as an approximate indicator to study interaction-induced transition.

%%%  FIG 8  %%%
\begin{figure}
    \centering
    \includegraphics[width=\linewidth]{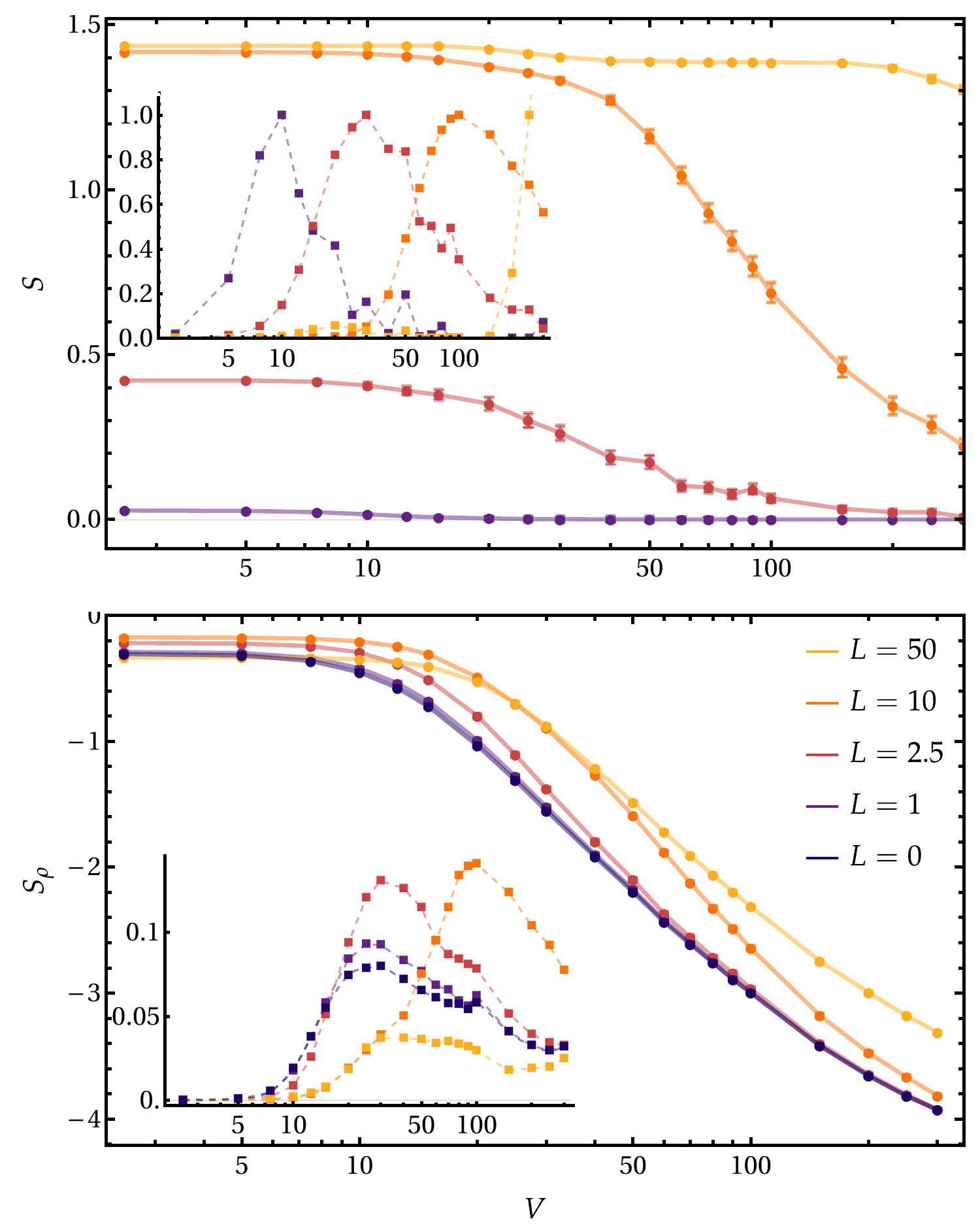}
    \caption{Single-particle occupation entropy (top panel) and position-space information entropy (bottom panel) as a function of $V$ for various interaction strengths $L$ averaged over 500 realizations. The two insets represent the variance associated to the entropies of their respective panel.
    \label{fig:fig_8}}
\end{figure}
%%% %%% %%% %%%

To conclude this study of this many-body localization model, the complementary point of view is considered, \ie the influence of interactions on the critical disorder strength $V_c$.
Analogously to the previous discussion on $L_c$, the mean of the single-particle occupation entropy as well as its variance are investigated.
They are plotted in the top panel of Fig.~\ref{fig:fig_8} and its inset, respectively.
Note that the variances in the inset have been rescaled to be all visible on the same plot.
For any value of $L$, $S$ tends to $0$ in the large $V$ limit, \ie a doubly occupied orbital.
As can be seen in the inset, the variance also exhibits a peak structure along the $V$ axis.
Note that for $L=50$, one can only see the onset of the decrease of $S$ as well as the onset of the peak of the variance.
The position of these maxima are used to define a critical disorder strength $V_c$ in presence of interactions.
The effect of interactions on these maxima is to delay them which means that the disorder-induced transitions are happening for larger disorder strengths.

The case of $L=50$ is particularly interesting as its variance present an additional smaller maximum for $V \approx 25$.
Indeed, the associated mean value of $S$ reaches a plateau equal to $2\ln(2)$ around this value of $V$ and then start to decrease towards $0$ at $V\approx200$.
This additional maximum corresponds to the Anderson localization of both particles in their respective side of the box observed in the last column of Fig.~\ref{fig:fig_6}.
Note that the convergence of the $L=1$ and $L=2.5$ curves is really slow due to some large outliers.
The cause of appearance of these outliers is explained in the \SupMat.

Once again the performance of the position-space information entropy as an approximate indicator is compared to the single-particle occupation entropy.
The mean values of $S_\rho(V)$ are plotted in the bottom panel of Fig.~\ref{fig:fig_8} and the associated variances are displayed in the inset.
The position and ordering of the maxima of the variance $\sigma(S_\rho)$ are in good agreement with the ones of the single-particle occupation entropy.
In addition, the variance of $S_\rho(V)$ for $L=50$ also shows a first small maxima and the onset of a second maxima similarly to what has been observed for the variance of $\sigma(S)$.

%%%%%%%%%%%%%%%%%%%%%%%%%%%%%%%%%%%%%%%%%%%%%%%%%%
\subsection{Interplay of interaction and disorder: DFT approximations}
\label{sec:interplay_approx}
%%%%%%%%%%%%%%%%%%%%%%%%%%%%%%%%%%%%%%%%%%%%%%%%%%

This final subsection deals with DFT approximations and their description of Mott--Anderson physics.
The focus will be on the qualitative description of the densities.
Indeed, because this is only a toy model of many-body localization quantifying the errors of approximate indicators is not really relevant in this context.
In addition, computing approximate statistic values would not give more physical insights than what has been investigated in the exact case of Sec.~\ref{sec:interplay_exact}.
Hence, the ground-state densities of the three KS DFT approximations considered in Sec.~\ref{sec:mott} are now studied in presence of disorder.

The first column of Fig.~\ref{fig:fig_6} shows that the three approximations perform similarly in the weak interactions regime.
The approximate densities are slightly over-delocalized in the weak and medium disorder regime while this error is getting smaller when $V$ is increasing.
This is expected as when $V$ grows the system is becoming more and more driven by the one-body part of the Hamiltonian, therefore reducing the importance of the approximate methods errors in the description of interactions.

As what has been observed for the $V=0$ case in Sec.~\ref{sec:mott}, LDA and EXX are unable to describe the strong interaction regime.
For weak disorder, the LDA and EXX wrongly predict delocalized densities (see $V=10/L=50$ panel) while in the large disorder strength limit, the two approximations display densities with a large number of peaks instead of the two sharp peaks of the exact solutions (see $V=75/L=50$ panel).
On the other hand, the SCE approximation describes qualitatively well the interplay of interaction and disorder.
Still there is a tendency to not localize enough analogous to what has been observed in the Mott transition without disordered potential.
This under-localization can be witnessed in the panels $L=10/V=10$ and $V=25$ for example.
This is due to the SCE localization of electrons happening too late (in terms of interaction strengths $L$) compared to the exact case (see discussion in Sec.~\ref{sec:mott}).
It is interesting to note that the external potential can improve the performance of SCE.
For example in the $L=10$ case, for no disorder (see Fig.~\ref{fig:fig_1}) or weak disorder ($V=10$ panel) SCE drastically fails to reproduce the two peak structure. 
However, for medium and strong disorder SCE correctly predicts the two peak structure observed in the exact case.

%%%  FIG 9  %%%
\begin{figure*}
    \centering
    \includegraphics[width=\linewidth]{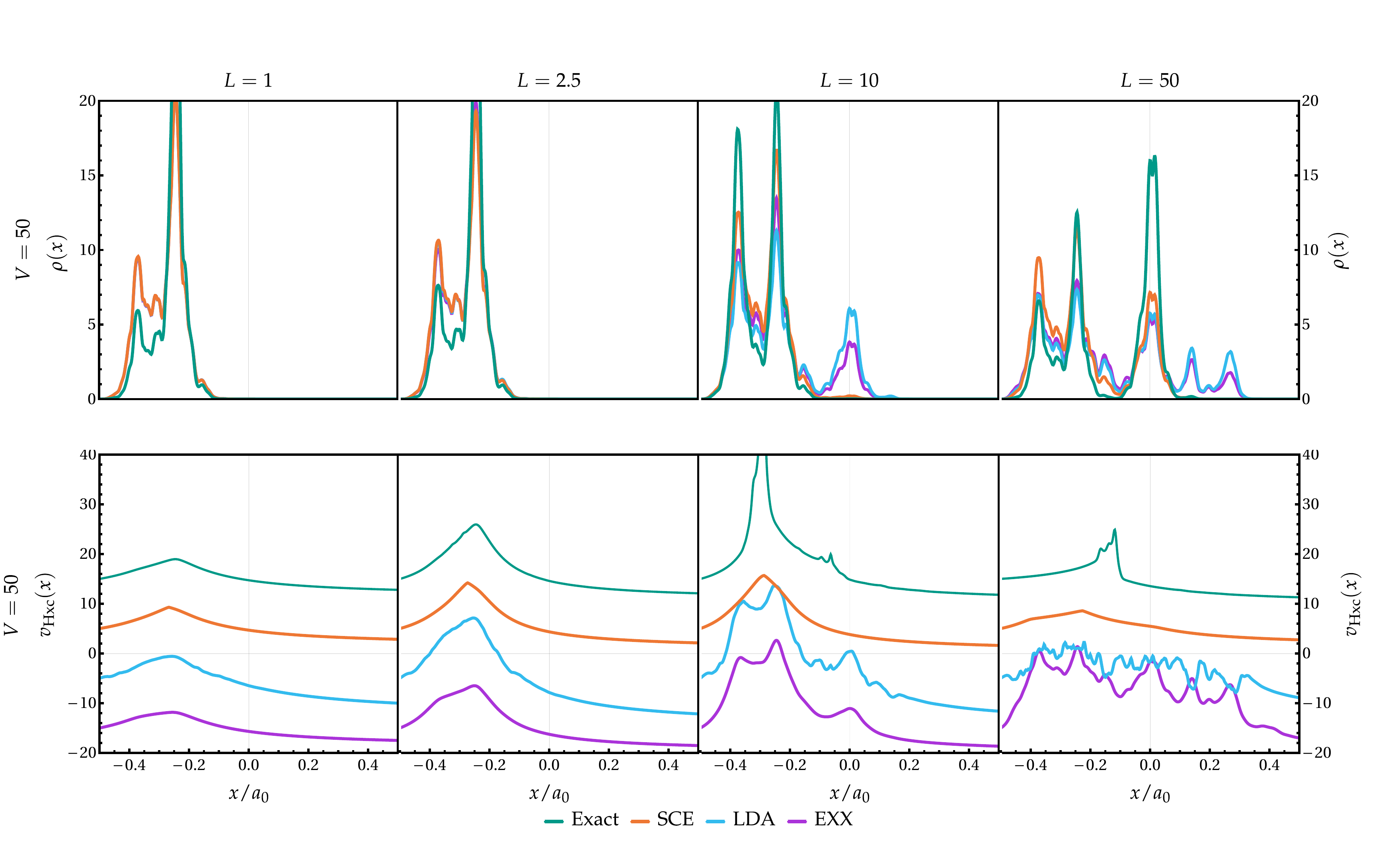}
    \caption{Exact, KS-SCE, KS-LDA and KS-EXX ground-state Hartree-Exchange-Correlation potentials for two electrons in the presence of a random potential ($V=50$) for various effective interaction strength $L$. We have added different constant shifts to each potential to make it easier to visualise the different shapes. 
    \label{fig:fig_9}}
\end{figure*}
%%% %%% %%% %%%

To conclude this result section, the KS potentials associated to the densities of Fig.~\ref{fig:fig_6} are shown in Fig.~\ref{fig:fig_9}. Notice that we have added different constant shifts to each potential to make it easier to visualise the different shapes. 
We see that when the interaction is weak ($L=1$ and $L=2.5$), all potential are qualitatively similar, with a maximum inside the region in which the density is large. When the density starts to show localization on well separated regions, the LDA and EXX starts to be qualitatively wrong: instead of having a maximum in the region in between the localization peaks of the density, they have a maximum where the density is localized. This is very similar to the error shown in stretched bonds in chemistry. KS SCE has a maximum localised in the right positions, but, again, way too low as it missed the kinetic correlation part. In the $L=50$ case, the exact potential shows a peak and a step structure, again reminiscent of what is observed in stretched heteronuclear bonds \cite{BuiBaeSni-PRA-89,GriBae-PRA-96,TemMarMai-JCTC-09,Helbig_2009,HodRamGod-PRB-16,Kohut_2016,Hodgson_2017}. KS LDA and EXX are completely wrong as, again, they display peaks where the density is localised, which make the self-consistent density ends up being too delocalized. KS SCE, although very different than LDA and EXX, is also not able to fully produce the correct step, in agreement with the analysis of Ref.~\onlinecite{GiaVucGor-JCTC-18}.

%=================================================================%
\section{Conclusions}
\label{sec:conclusion}
%=================================================================%

Many-body localization, the field of physics studying isolated many-body systems in the presence of disorder, is an increasingly active research domain. \cite{Abanin_2017,Alet_2018}
This effervescence is due to its relevance for a large panel of areas of physics as well as its connections to the foundation of statistical mechanics. 
The main advances in understanding this complex phenomenon came from lattice Hamiltonian models, and especially from the one-dimensional ones.
Indeed, these models offers an incredible gain in terms of computational cost while retaining most of the physics of their real-space counterparts.
Yet, a clear understanding of the differences between real-space and lattice models remains desirable to be aware of possible flaws of lattice models.

The one-dimensional Hubbard-Anderson Hamiltonian
\begin{equation}
    \label{eq:HubbardAnderson}
    \hat{H} =  -t \sum_{ij,\sigma}\mleft(\hat{a}_{i\sigma}^\dagger\hat{a}_{j\sigma} + \hat{a}_{j\sigma}^\dagger\hat{a}_{i\sigma}\mright) + U \sum_i\hat{n}_{i\uparrow}\hat{n}_{i\downarrow} + \sum_{i,\sigma} V_i\hat{n}_{i\sigma},
\end{equation}
has been widely used as a model for interacting electrons in the presence of disorder.
The aim of this work is to investigate the real-space analog of this model described in Eqs.~(\ref{eq:hamiltonian}-\ref{eq:random_potential}).
One major difference between these two systems is the long-range character of the Coulomb-like interaction of Eq.~\eqref{eq:interaction}.
On the other hand, in the Hubbard-Anderson model, electrons are restricted to interact with each other only when they are on the same site (short-range interactions).
We should mention that inclusion of long-range interactions in many-body localization lattice models is possible and has been investigated by some groups recently. \cite{Burin_2015,Hauke_2015,Nandkishore_2017,Nag_2019,Vu_2022}

While many-body localization should happen for both ground and excited states, this work has considered only the first ones because of computational restrictions due to the real-space nature of our system.
For the same reason, only two electron systems have been considered.
Yet, this simple system still carries much valuable information about the interplay of long-range interactions and disordered potentials.
In addition, this system allowed to gauge the performance of localization indicators usable in real-space models because the entanglement measures widespread in lattice Hamiltonians can not be straightforwardly transferred to real space.
Namely, the single-particle occupation entropy and the position-space information entropy, relying respectively on the one-body RDM and the electronic density, have been considered.
While it has been observed that the former is a more reliable indicator to study localization transition induced by interaction, the latter has the practical advantage of being usable for DFT approximations.

The study of the numerically accurate many-body solutions of this Hamiltonian allowed to observe some trends about the interplay of disorder and interactions.
We showed that increasing the disorder strength delays the Mott-like transition to larger interaction strength as well as making transitions much steeper.
The other point of view, namely the influence of interactions on disorder-induced localization has been investigated as well.
In this case, it has been showed that the disorder-induced localization is happening at larger disorder strength for large interaction strength.
In addition, in this large interaction regime, another transition has been detected in which the electrons stay in singly occupied orbital but these orbitals become localized in space.

The model also allowed us to analyse in depth the performance of different density functional approximations. Our results show that the exact KS potential needs to have features very similar to those that have emerged from the study of the KS potential for molecular systems, in particular peaks and steps. The failure of (semi)local approximations, and also of exact exchange, which are known to miss these features, is in this case particularly spectacular, as it leads to completely delocalized densities when the exact ones are localised. It could be seen as an extreme case of the delocalization error \cite{BryAdeDalJoh-WIREs-22}. The SCE functional, although performing qualitatively much better, has still notable failures because it misses the kinetic correlation part. This study suggests that this simple model system of electrons in a box in the presence of disorder could be used as a severe test for new DFT approximations. 

%=================================================================%
\section*{Supplementary material}
\label{sec:suppmat}
%=================================================================%

%%%%%%%%%%%%%%%%%%%%%%
\begin{acknowledgements}
%%%%%%%%%%%%%%%%%%%%%%
This work was supported by the Netherlands Organisation for Scientific Research (NWO) under Vici grant 724.017.001 and by the H2020/MSCA-IF ``SCP-Disorder'' [grant 797247].

Juri Grossi whishes to acknowledge the U.S. Department of Energy, National Nuclear Security Administration, Minority Serving Institution Partnership Program, under Award DE-NA0003866.
%%%%%%%%%%%%%%%%%%%%%%
\end{acknowledgements}
%%%%%%%%%%%%%%%%%%%%%%

%%%%%%%%%%%%%%%%%%%%%%
\bibliography{MBL_KSSCE,bib_clean.bib}
%%%%%%%%%%%%%%%%%%%%%%

\end{document}

\title{Supplementary information of "Real space Mott-Anderson electron localization with long-range interactions: exact and approximate descriptions"}

\author{Antoine \surname{Marie}}
\affiliation{\VU}
\author{Derk \surname{Kooi}}
\affiliation{\VU}
\author{Juri \surname{Grossi}}
\affiliation{\UCM}
\author{Michael \surname{Seidl}}
\affiliation{\VU}
\author{Ziad H. \surname{Musslimani}}
\affiliation{\FSU}
\author{Klaas \surname{Giesbertz}}
\affiliation{\VU}
\author{Paola \surname{Gori-Giorgi}}
\affiliation{\VU}

\maketitle

\begin{figure*}
    \centering
    \includegraphics[width=0.45\linewidth]{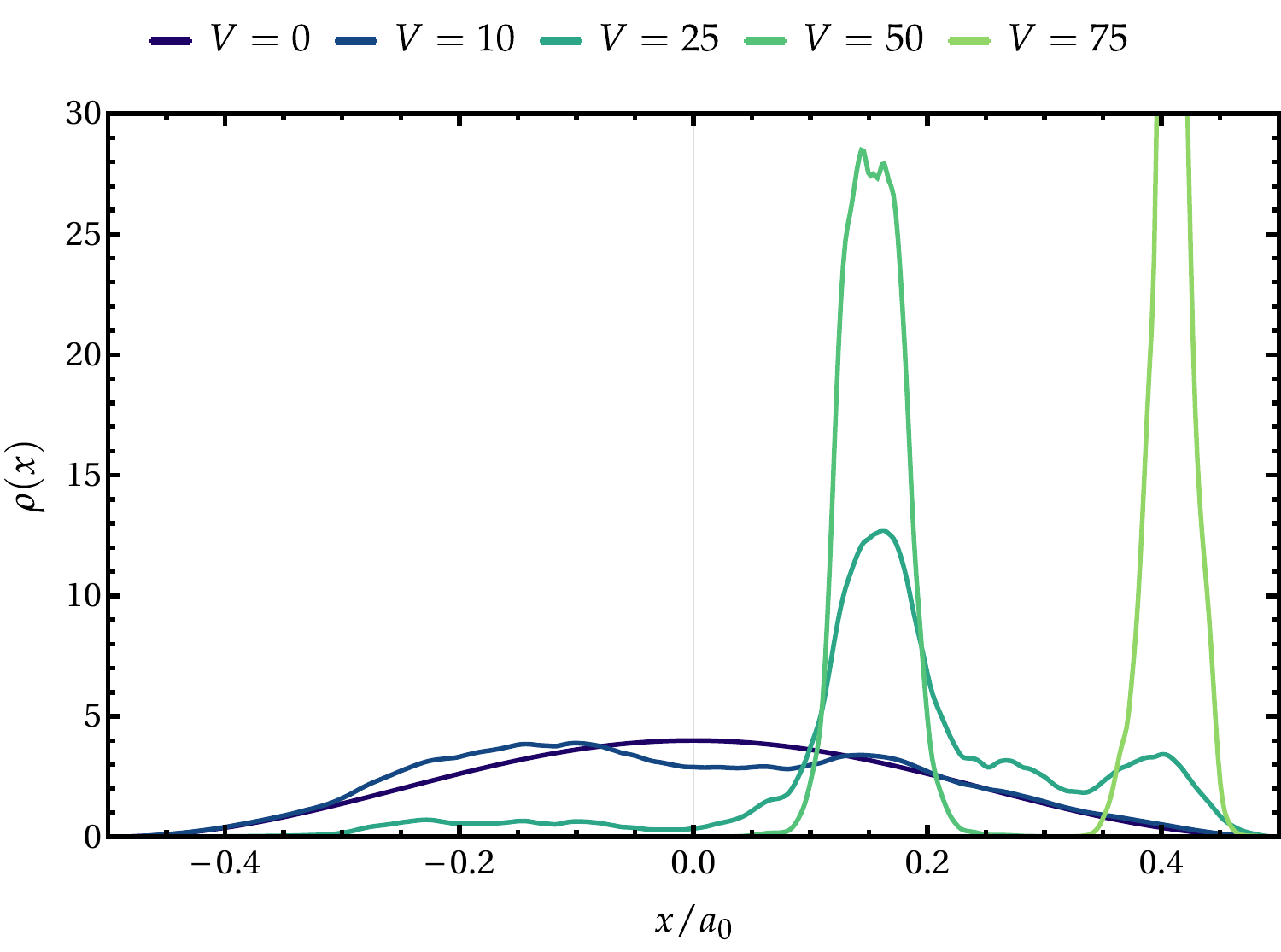}
    \includegraphics[width=0.45\linewidth]{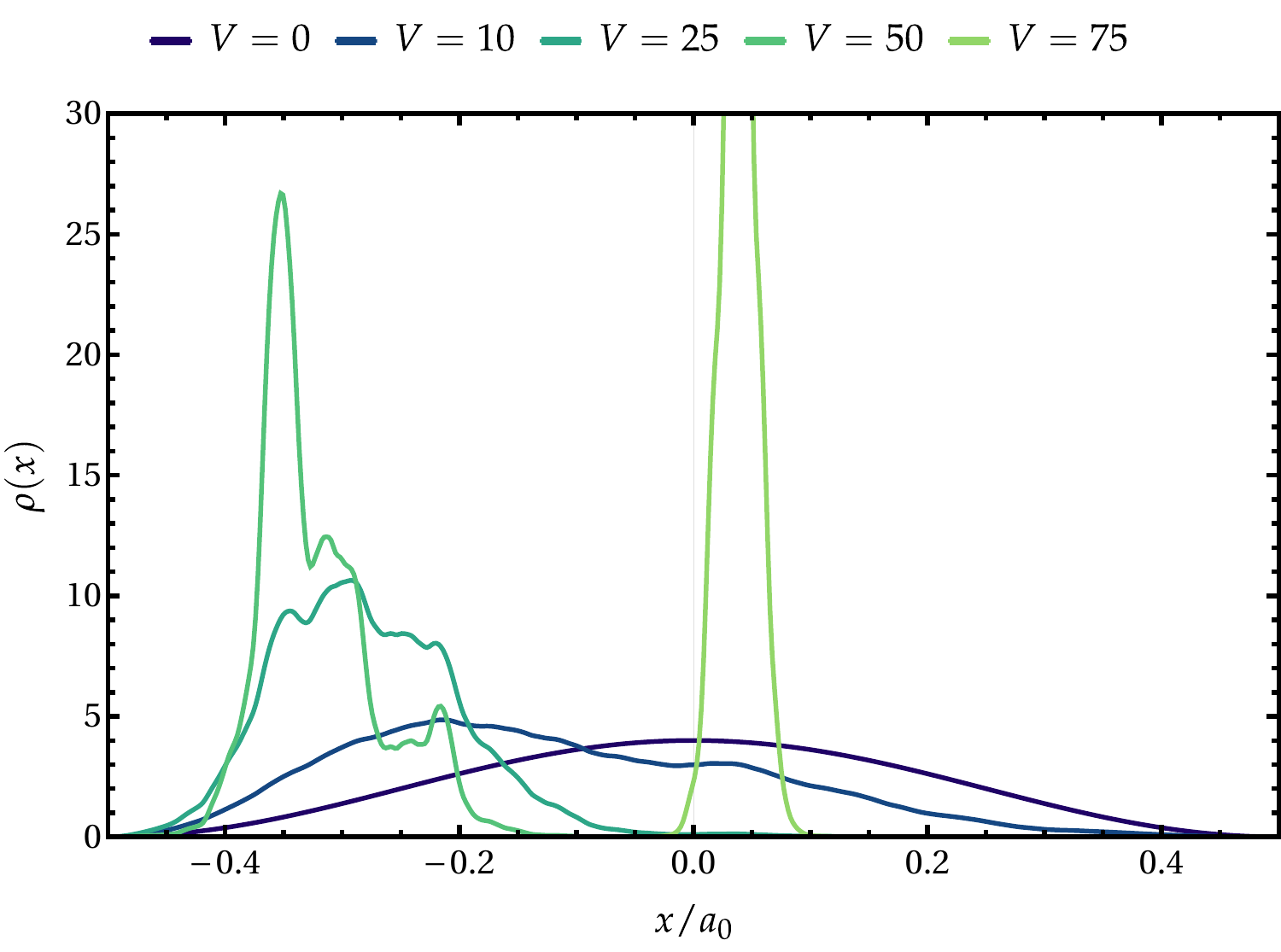}
    \includegraphics[width=0.45\linewidth]{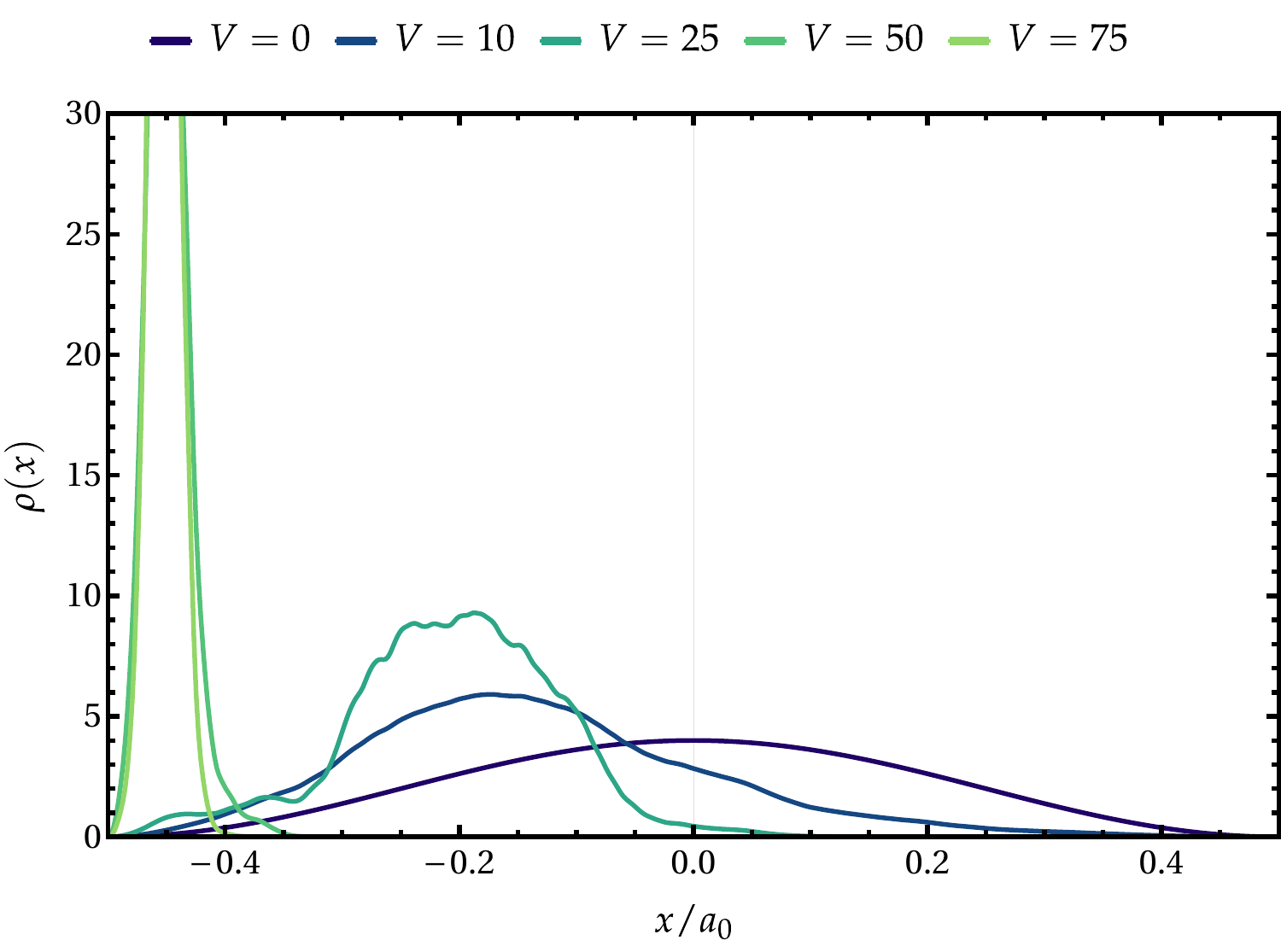}
    \includegraphics[width=0.45\linewidth]{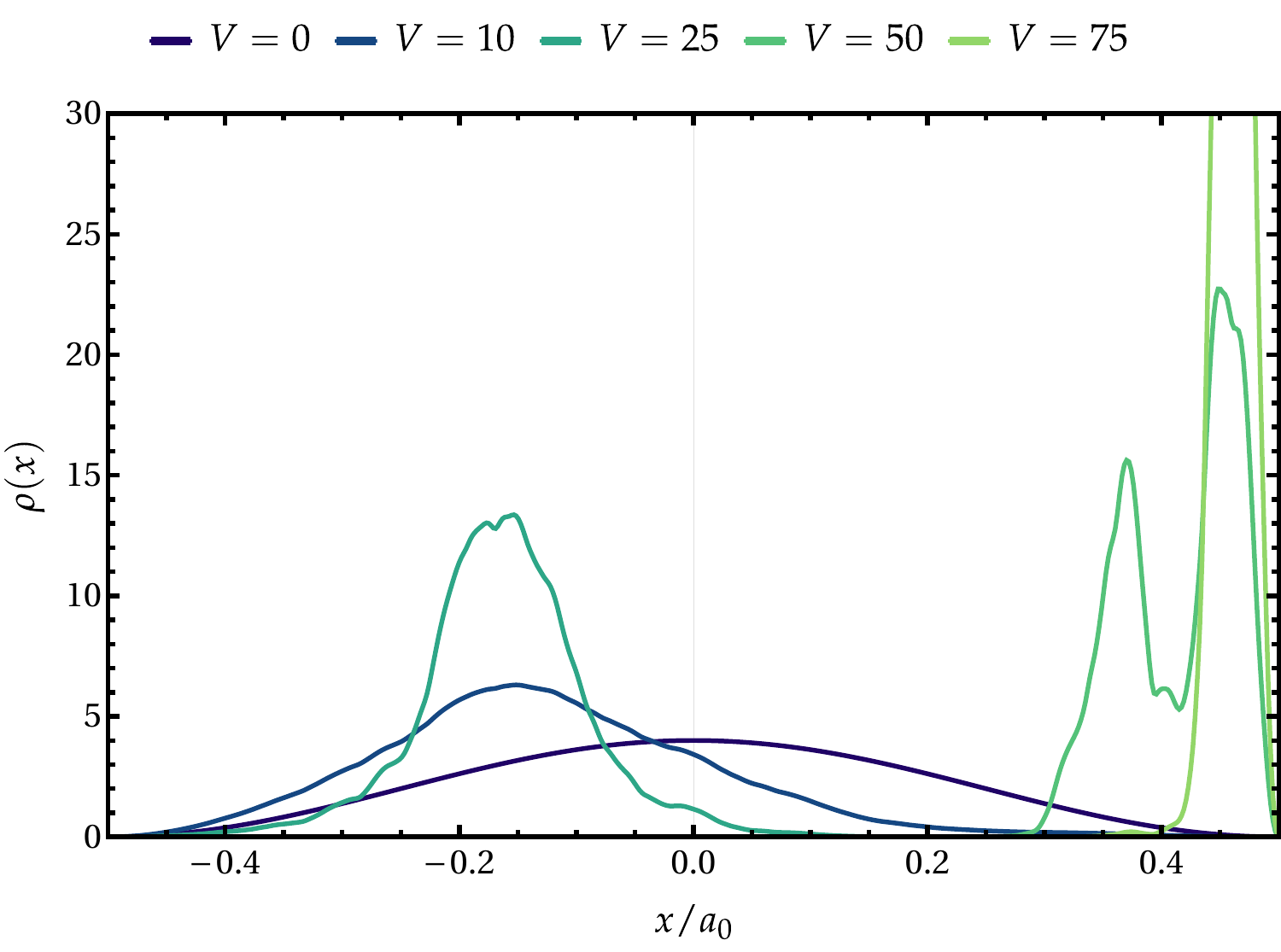}
    \caption{Electronic densities (solid lines) of two non-interacting electrons for various values of the potential strength $V$ and four different realizations of the disorder. The height of the potential (dashed line) is non-indicative, it has been adjusted just to show its shape on the same plot.
    \label{fig:fig_supmat1}}
\end{figure*}

Figure 4 of the manuscript has been reproduced in Figure~\ref{fig:fig_supmat1} with four different realizations of disorder.
The qualitative evolution of the densities is similar to the realization discussed in the main manuscript.
These additional examples show that the localization can happen at different positions for various values of the disorder strength within the same realization.
The transition between peaks localized at two different positions results in an avoided crossing.
Near the avoided crossing, the ground state is a mix of the two states leading to a two-peak structure (see bottom-right panel) even if the two particles are not interacting.
These avoided crossings are responsible of the slow convergence of the mean of the single-particle occupation entropy in the weak interaction regime ($L=1$ and $L=2.5$).
Indeed, at an avoided crossing two orbitals are occupied which gives a value of $S$ close to $0.7$ which slows the convergence of the mean of $S$ towards values close to $0$.
These outliers have been removed when computing the mean values displayed in Figure~8 of the main manuscript to speed up the convergence of the mean.

%%%  FIG 1  %%%
%%% %%% %%% %%%

\begin{figure*}
    \centering
    \includegraphics[width=\linewidth]{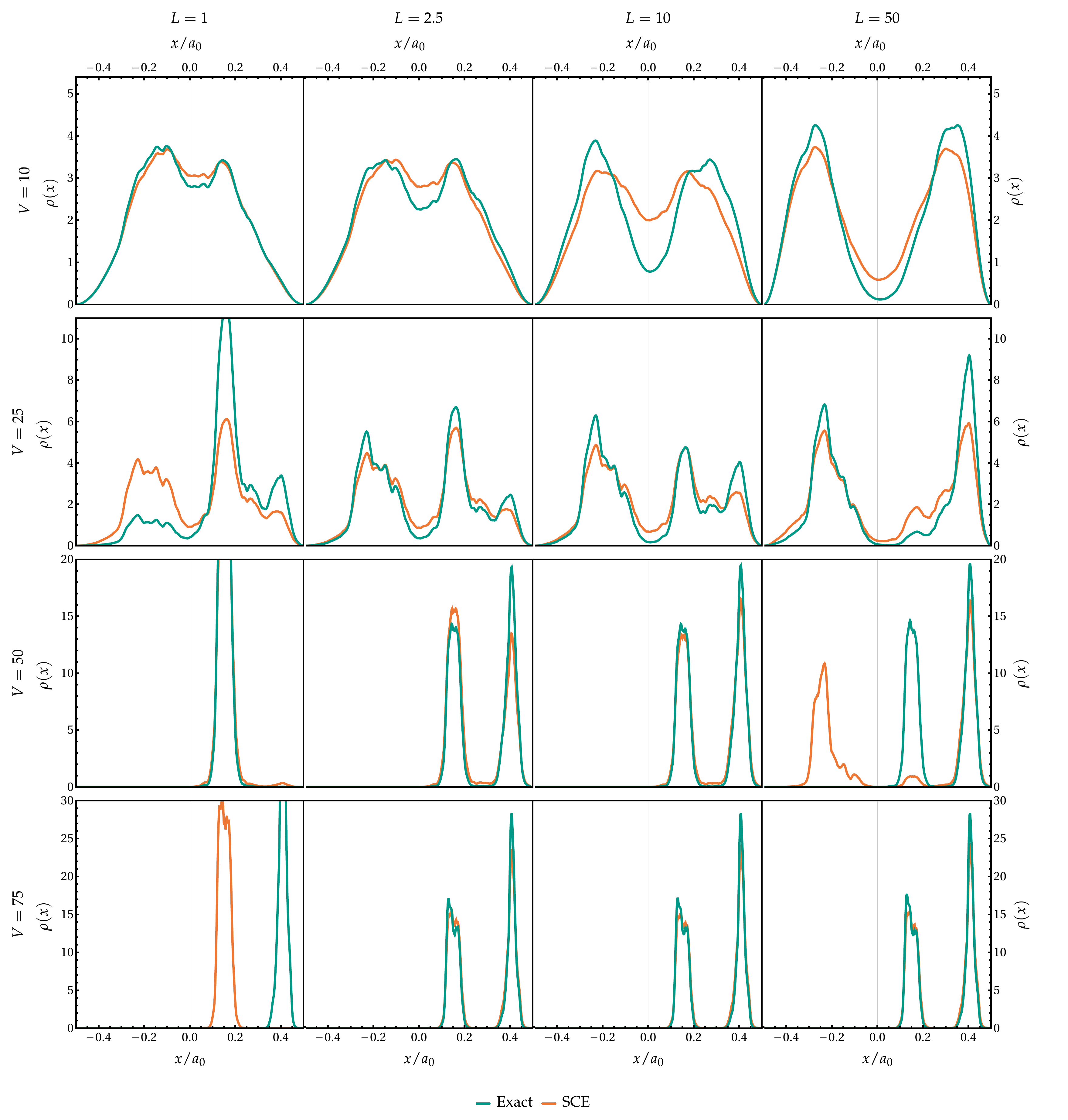}
    \caption{Exact and KS SCE ground-state electronic densities for a given realization of disorder with various effective interaction strengths $L$ and disorder strengths $V$.
    \label{fig:fig_supmat2}}
\end{figure*}

In addition, we also report the exact and KS SCE ground-state densities in the presence of interactions for two other realizations than the one displayed in Figure~6 of the main manuscript. 
Note that these two realizations are the same as the ones in the top row of Figure~\ref{fig:fig_supmat1}.
Once again the qualitative analysis of the evolution of the densities is the same as the one discussed in the main manuscript.
One can see on these two additional examples that the avoided crossings at the KS SCE level do not happen at the same values of $V$ as the exact ones. 
This leads to differences of position of localization between KS SCE and the exact results as can be observed in the $L=1/V=75$ and $L=50/V=50$ of Figure~\ref{fig:fig_supmat2}.

%%%  FIG 2  %%%
%%% %%% %%% %%%

%%%  FIG 3 %%%
\begin{figure*}
    \centering
    \includegraphics[width=\linewidth]{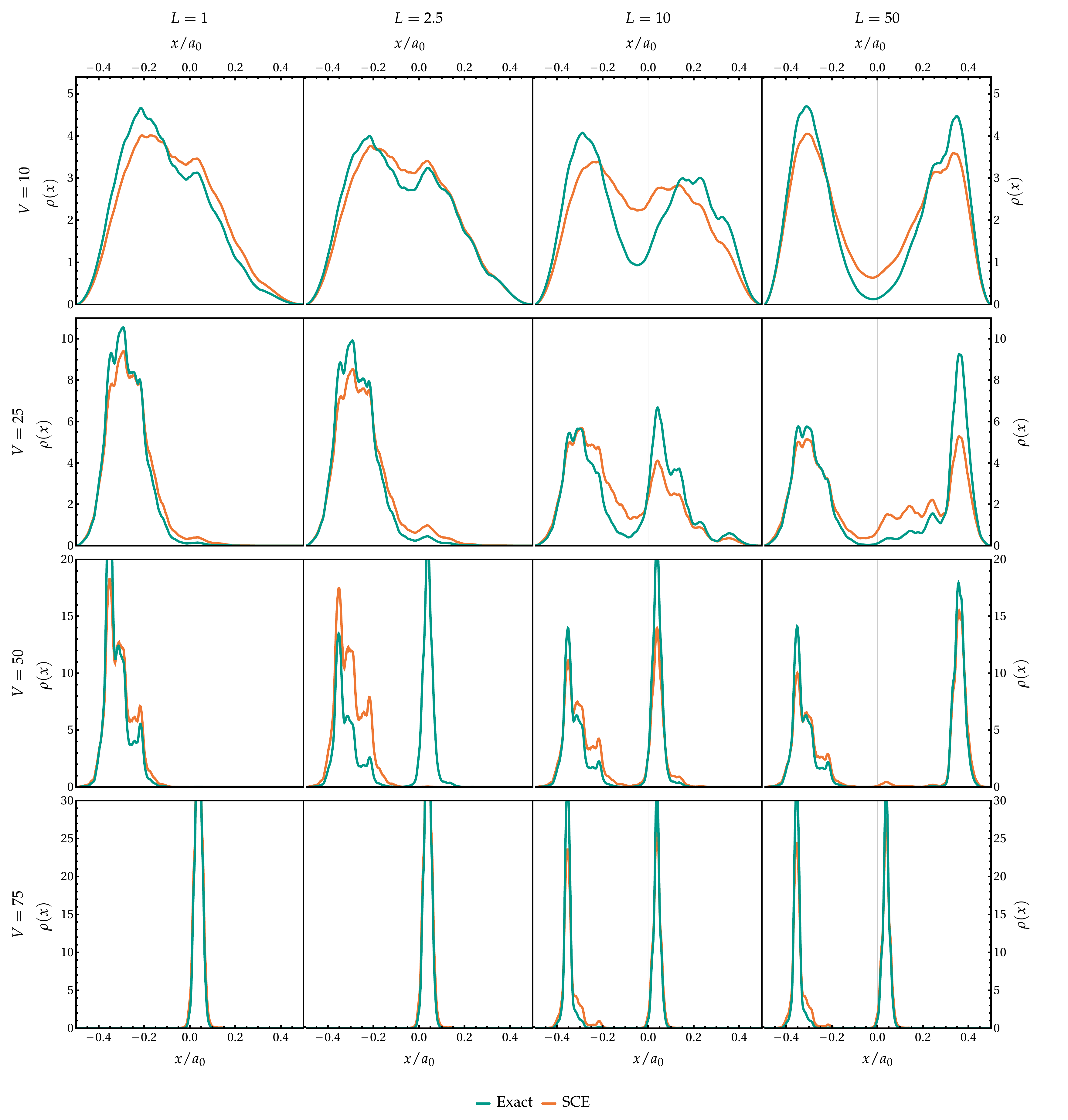}
    \caption{Exact and KS SCE ground-state electronic densities for a given realization of disorder with various effective interaction strengths $L$ and disorder strengths $V$.
    \label{fig:fig_supmat3}}
\end{figure*}
%%% %%% %%% %%%

Finally, the bottom panel of the seventh figure of the main manuscript is reproduced here without shift to show the real values of the position-space information entropy.

%%%  FIG 4  %%%
\begin{figure}
    \centering
    \includegraphics[width=\linewidth]{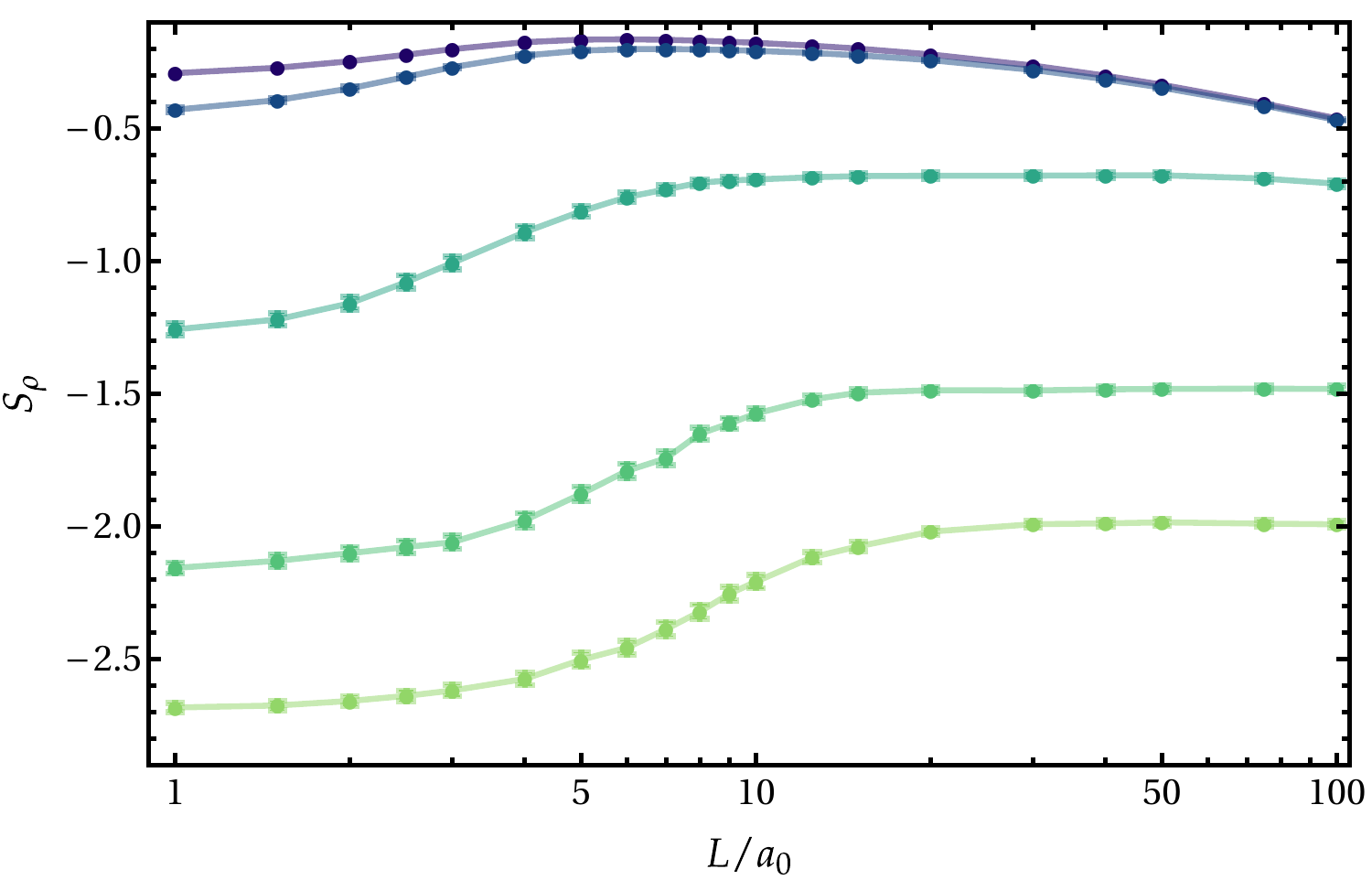}
    \caption{Position-space information entropy as a function of $L$ for various disorder strengths $V$ averaged over 200 realizations.
    \label{fig:fig_supmat4}}
\end{figure}
%%% %%% %%% %%%